\documentclass[english,aps,prb,twocolumn]{revtex4-1}
\usepackage{color}
\usepackage[english]{babel}
\usepackage{natbib}
\usepackage{amsmath}
\usepackage{amssymb}
\usepackage{graphicx}
\usepackage{indentfirst}

\makeatletter

\renewcommand{\emph}[1]{\textit{#1}}

\begin{document}

\title{Two-electron selective coupling in an edge-state based conditional phase shifter}
\author{Laura Bellentani$^{1}$, Gaia Forghieri$^{2}$, Paolo Bordone$^{1,2}$ and Andrea Bertoni$^{1}$}
\affiliation{$^1$S3, Istituto Nanoscienze-CNR, Via Campi 213/A, 41125 Modena, Italy}
\affiliation{$^2$Dipartimento di Scienze Fisiche, Informatiche e Matematiche, Universit{\`a} degli Studi di Modena e Reggio Emilia, Via Campi 213/A, 41125 Modena, Italy}
\begin{abstract}
We investigate the effect of long-range Coulomb interaction on the two-electron scattering in the integer quantum Hall regime at bulk filling factor 2.  A parallel version of the Split-Step Fourier method evolves the exact two-particle wave function in a 2D potential background reproducing the effect of depleting gates in a realistic heterostructure, with the charge carrier represented by a localized wavepacket of edge states.  We compare the spatial shift induced by Coulomb repulsion in the final two-electron wave function for two indistinguishable electrons initialized in different configurations according to their Landau index, and analyze their bunching probability and the effect of screening.  We finally prove the feasibility of the present operating regime as a two-qubit conditional phase shifter to generate entanglement from product states. 
\end{abstract}


\maketitle

\section{Introduction}
Over the years, single-electron and two-electron interference have been realized in a large variety of devices operating in the Integer quantum Hall regime, making them a ideal platform for electron quantum optics\cite{GrenierMPLB2011,RousselPSSB2017,GlattliPSSB2017,LOCANENJP2019,RodriguezPhD2019} and a possible candidate for quantum computing architectures based on flying qubits\cite{JINATURE2003, NEDERPRL2006, KANGPRB2007, LepagePRA2020}. 
Early implementations of the electronic Mach-Zehnder interferometer (MZI) at bulk filling factor one prove self-interference of the electron wavefunction and the viability of coherent transport in edge states\cite{ROULLEAUPRL2008, DEVIATOPRB2011, BEGGIJOPCM2015}, but are affected by a fundamental geometrical limit that jeopardizes their concatenation in series. 
A new geometry has been then proposed recently by Giovannetti et al. \cite{GIOVANNETTIPRB2008}, where the scattering between the first two copropagating edge channels is generated by an alternative design of the beam splitter\cite{KARMAKARPRB2015, BELLENTANIPRB2018}. 
\begin{figure*}[tb]
\centering
\begin{minipage}[c]{0.51\textwidth}
\includegraphics[width=1\textwidth]{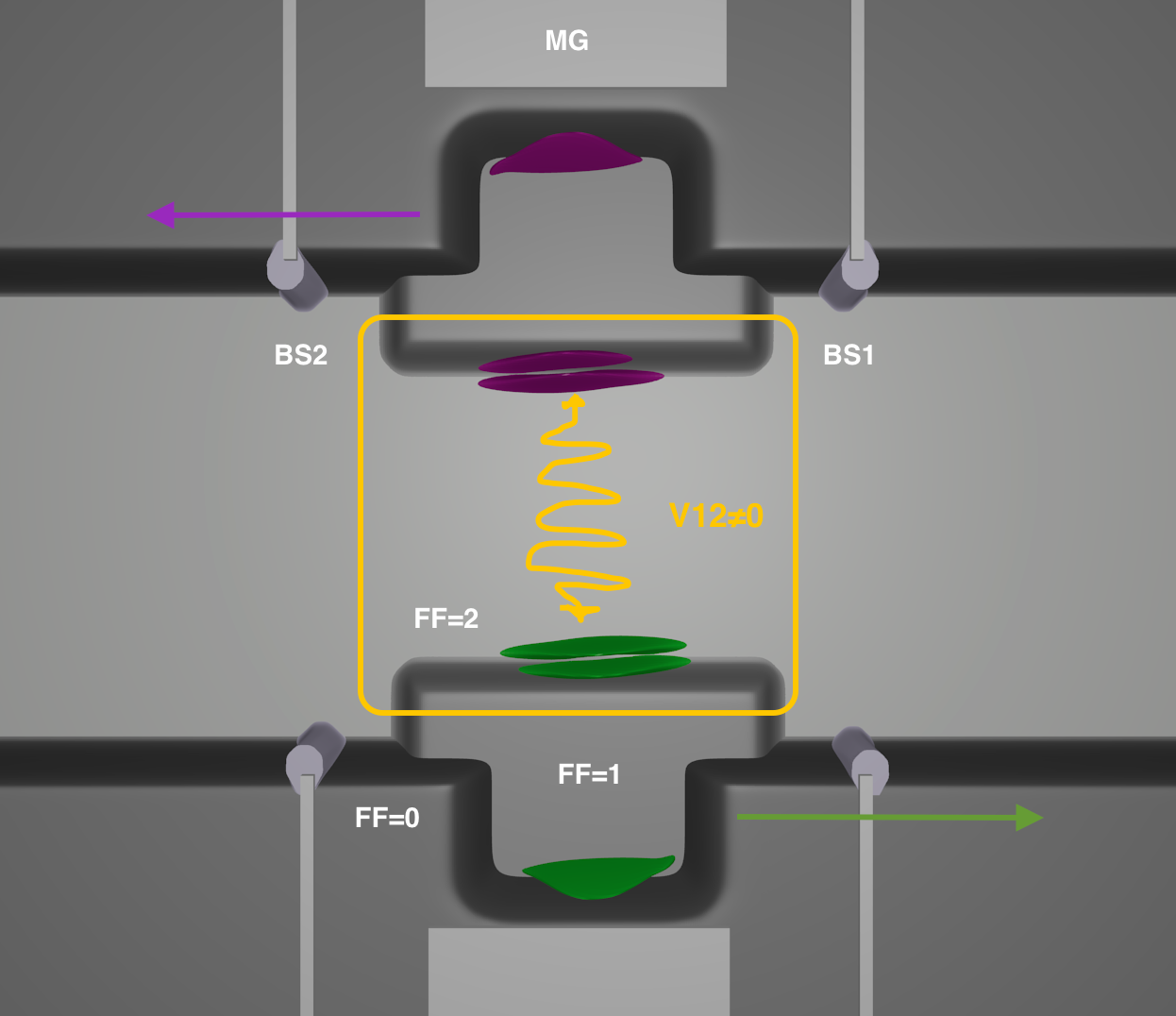}
\end{minipage}
\begin{minipage}[c]{0.48\textwidth}
\includegraphics[width=1\textwidth]{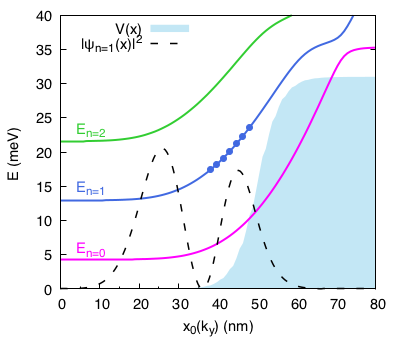}
\end{minipage}
\caption{(a) Two parallel multichannel MZIs, as in Ref.~\onlinecite{BELLENTANIPRB2018}, define a conditional phase shifter ${\bf{T}}$ in the IQH regime.  The two channels with Landau index $n=1$ are localized at the inner edge of the mesa, where the filling factor (FF) is two.  The yellow squared box identifies the active region, where electrons interact.  The channels with $n=0$ are localized at the outer edge of the mesa and can be further separated by increasing the width of the region at FF=1.  The purple/green wavepackets show the single-electron density probability of the electrons in the two available channels, at opposite sides of the 2DEG.  (b) Bandstructure of the active region of the conditional phase shifter in panel (a) at one of the two edges of the device.  The first 3 Landau levels ($n=0,1,2$) are displayed as solid lines labelled with $E_n$; the dots at $n=1$ identify the edge states combined to generate a Gaussian wavepacket in the second channel with $\sigma=40$~nm and a central energy $E^0=21$~meV.  The single-electron probability distribution of the electron wave packet in the transverse direction of the second edge channel is also displayed (dashed line). }
\label{fig:realdev}
\end{figure*}
The scalability of this new approach allows, in principle, the implementation of two-qubit logic gates, as the Hanbury-Brown-Twiss interferometer \cite{OLIVERSCIENCE1999, BUTTIKERPHYSICAE2003, CHUNGPRB2005, SAMUELSSONPRL2004, NEDERNATURE2007, GLATTLISCIENCE2016}, where exchange symmetry induces the Hong-Ou-Mandel (HOM) effect \cite{BELLENTANIPRB2019, Ferraro2018_EuPJST, MARGUERITEPRB2016, MARIANJPHYSCMAT2015, WAHLPRL2014}, and the Conditional Phase Shifter (CPS), for entanglement generation\cite{Bertoni_PRL2000}. 
Together with the effect of electron-electron interaction, the interplay between the geometry of the device and the electron correlations plays a crucial role in the quantum logic gate operation\cite{BORDONESST2019}. 

In the literature, the numerical simulation of Hall interferometers usually exploits the chirality of edge states to model electron transport in effective 1D schemes\cite{KANGPRB2007, CHIROLLIPRL2013}.
Moreover, delocalised edge states are often considered as current-carrying states.
Differently, we simulate single and two-electron transport in a full-scale 2D Hall nanodevice by using a time-dependent framework based on the Split-Step Fourier method\cite{Kramer2010_PS}, where electrons are described by single-charge wavepackets. 
We numerically setup the 2D potential landscape generated by modulation gates in order to compute the edge states with the exact shape induced by our design of the confining barrier; such states are then linearly combined with a Gaussian weight function.
With a large but affordable computational cost\cite{BORDONESST2019}, this method provides access to the dynamical properties of an interacting system of electrons directly from the exact two-particle state and allows us to introduce in a rigorous way electron-electron repulsion.
This proved to be relevant in devices whose functioning is based on two-electron scattering, as the HOM interferometer in Ref.[\cite{BELLENTANIPRB2019}], where we observe the transition from an exchange-driven to a Coulomb-driven bunching of strongly-localized wavepackets. 
Moreover, by encoding the initial electron state in a Gaussian wavepacket\cite{BEGGIJOPCM2015, BELLENTANIPRB2018, BELLENTANIPRB2019}, we reproduce the injection of an \textit{hot electron} by means of single-electron sources\cite{BUTTIKERPLA1993, MAHEPRB2010, BOCQUILLONSCIENCE2013, BLUMENTHALNATPHYS201307,  DUBOISNATURE2013, KATAOKAPSS(2017), Bauerle_RPP2018}, as recently proposed theoretically by Riu et al. with quantum dot pumps\cite{RYUPRL2016}. 

In this paper, our full-scale numerical approach is applied to simulate two-electron scattering in the active region of a solid-state conditional phase shifter\cite{BORDONESST2019}, as the one depicted in Fig.~\ref{fig:realdev}(a). 
The figure shows the interferometer pattern created by the external potential (grey structure) induced by modulation gates and the electron wavepackets at an intermediate time. 
Two multichannel MZIs\cite{GIOVANNETTIPRB2008, BELLENTANIPRB2018} in the IQH regime are concatenated in parallel, and generate four channels, two of them (ground and first excited) running at each side of the 2DEG. 
Self-interference in each MZI\cite{BELLENTANIPRB2018} is affected by a selective Coulomb interaction that couples only those electron states localized in the first excited channels. 
Here the distance between the two electron paths is decreased by a potential mesa, where the bulk filling factor is 1, so that their mutual interaction is increased. 
Figure~\ref{fig:realdev}(b) shows the bandstructure of the active region for Coulomb coupling (yellow box). 
Here, the spatial confinement of the transverse probability distribution for the single-electron wave function (dashed black line) ensures the absence of tunneling between counterpropagating channels.

According to its strength, Coulomb coupling affects the two-electron state at the output of the device, by selectively rotating only the component of the wave function with both electrons in the first excited states. 
This realizes the transformation\cite{BORDONESST2019, Bertoni_PRL2000}
\begin{equation}  \label{eq:tmatrix}
{\bf{T}}(\gamma )=\left(\begin{array}{cccc}
1 & 0 & 0 & 0 \\
0 & 1 & 0 & 0 \\
0 & 0 & 1 & 0 \\
0 & 0 & 0 & e^{i\gamma} 
\end{array}\right).
\end{equation}
The present device provides the ideal playground for our numerical model: the geometrical parameters of the potential landscape, e.g. the length of the coupling region and the distance between the channels, affect significantly the strength of Coulomb repulsion and therefore the corresponding angle $\gamma$. 

This article is organized as follows. 
In Sec.~\ref{sec:physsys} we summarize the numerical model of localized carriers in edge states in the Integer Quantum Hall regime and our simulation approach of the two-electron device. 
Then, Sec.\ref{sec:res}~A describes the selective action of the Coulomb interaction in a simple geometry for the active region, with two sharp potential barriers at the edges of the 2DEG. 
We analyze how the Landau level bandstructure affects the effectiveness of electron-electron repulsion, and calculate the total amount of energy exchanged during the two-electron scattering in Sec.\ref{sec:res}~B. 
After computing the bunching probability for short interchannel distances, we include screening effects on the two-particle dynamics in Sec.\ref{sec:res}~C. 
Finally, we adopt a more realistic model of the confining potential with smoothed Fermi-like barriers, and in Sec.\ref{sec:res}~D we predict the $\gamma$ factor expected in a full-scale conditional phase shifter.

\section{Physical system and Numerical model}\label{sec:physsys}
In our numerical simulations, two interacting electrons with charge $q=-e$ and an effective mass $m^*$ propagate in a confined 2DEG on the $xy$-plane and are immersed in a perpendicular magnetic field ${\bf B}=B\hat{z}$. 
The effect of the magnetic field on the electron transport is described in the Landau gauge, which introduces the vector potential ${\bf A}_i=Bx_i\hat{y}$ in the single-particle Hamiltonian $\hat{H}_i$, with $i=1,2$ indicating the first or second electron, respectively. 
In presence of Coulomb interaction $\hat{V}_{12}$ and a confining potential $\hat{V}$, the two-electron Hamiltonian reads:
\begin{eqnarray}\label{eq:H_effective}
\hat{H}&=&\frac{(\hat{p}_1-q\hat{A}_1)^2}{2m^*}+\frac{(\hat{p}_2-q\hat{A}_2)^2}{2m^*}+ \hat{V}_1+ \hat{V}_2+\hat{V}_{12} \nonumber \\
&= &\sum_{i=1,2} \left( \frac{\hat{p}_{x_i}^2}{2m^*}+\frac{\hat{p}_{y_i}^2}{2m^*}+ \frac{eB\hat{x}_i\hat{p}_{y_i}}{2m^*}+\frac{e^2B^2\hat{x}_i^2}{2m^*} + \hat{V}({\bf r}_i) \right)    \nonumber \\ &&+ \hat{V}_{12}({\bf r}_1,{\bf r}_2).\label{eq:H2p}
\end{eqnarray}
Here, magnetic components of the kinetc operators in the Landau gauge clearly couples the $\hat{x}$-coordinate in the real space and the $\hat{k}_y$-coordinate in the reciprocal space, for each particle. 

At the initial time, when the two counterpropagating electrons are distant, the mutual interaction is negligible ($V_{12}\simeq 0$) and each single-particle Hamiltonian $\hat{H}_i$ in Eq.~(\ref{eq:H2p}) is separable on the real-space domain. 
By adopting the \textit{ansatz} $\phi_i=\varphi_{n_i,k_i}(x_i)e^{ik_iy_i}$, the non-interacting Hamiltonian can be expressed in the effective 1D form
\begin{eqnarray}
\hat{H}_i\phi_i&=\big[-\frac{\hbar^2}{2m^*}\frac{\partial^2}{\partial x_i^2}+\frac{1}{2}m^*\omega_c^2(x_i-x^0_i)^2 \\
&+\hat{V}(x_i,y_i)\big]\varphi_{n_i,k_i}(x_i)e^{ik_iy_i},
\end{eqnarray} 
where $k_i$ is the wave vector in the $\hat{y}$-direction of the reciprocal space, $|\omega_c|=eB/m^*$ is the cyclotron frequency, $n_i$ is the Landau index and $x^0_i=-eBk_i/\hbar$ is the center of an effective parabolic confinement in the $\hat{x}$-direction induced by the magnetic field.

In the presence of a translationally invariant potential in the $\hat{x}$-direction, $V(x,y)=V(x)$, by selecting a wave vector $k$ and Landau index $n$, we identify the eigenstates $\varphi_{n,k}(x_i)$ that diagonalize the single-particle effective Hamiltonian in 1D:
\begin{equation}
\hat{H}^{eff}(x)=-\frac{1}{2m^*}\frac{\partial^2}{\partial x^2}+\frac{1}{2}m^*\omega_c^2(x-x^0)^2+\hat{V}(x).
\end{equation} 
The Hamiltonian above determines the localization of the electron wave function in the transverse direction of the device (orthogonal to the propagation direction), due to the presence of the magnetic confining potential, $V_B(x)=\frac{1}{2}m^*\omega_c^2(x-x^0)^2$ which adds to the external one, $V(x)$.

Note that in the bulk of the confined 2DEG, where $V(x)\simeq 0$, the eigenfunction $\varphi_{n,k}$ coincides with the eigenstate of an harmonic oscillator with frequency $\omega_c$. In presence of a non-negligible confining potential, the eigenstates of the effective Hamiltonian $\hat{H}^{eff}_i$ must be computed numerically, and their shape depends on the smoothness of $V(x)$. The single-particle wavefunction $\phi(x,y)=\varphi_{n,k}(x)e^{iky}$ is indeed called \textit{edge state}, and it is composed by a delocalized plane-wave term in the longitudinal direction, $e^{iky}$, and a confined wavefunction $\varphi_{n,k}(x)$ in the transverse one. The corresponding eigenenergy for a given value of the quantum numbers $n$ and $k$ is
\begin{equation}
E_{n}(k)=\hbar\omega_c(n+\frac{1}{2})+\epsilon_{n}(k),
\end{equation}
which depends on the wavevector $k$ only if $\epsilon_{n_i}(k_i)\neq 0$, i.e. in proximity to the confining edge barrier.
\begin{figure}[t]
\centering
\includegraphics[width=1.\columnwidth]{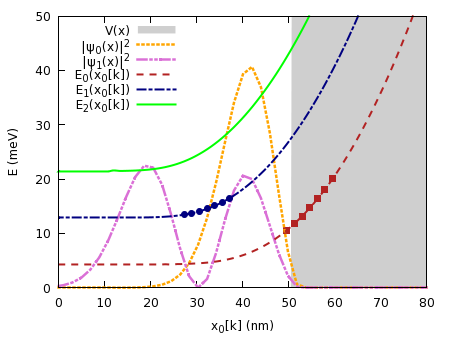}
\caption{Bandstructure of the active region induced by the sharp potential barrier (grey shaded area) for the first three edge channels at $B = 5$~T.  The dots define the centers $x_0(k)$ of the edge states contained in the Gaussian wavepackets with $n=0$ and $n=1$.  The transverse profiles of the probability density for a wavepacket injected with a central energy $E^0 = 15$~meV and $\sigma = 40$~nm in $n=0$ and $n=1$ are also displayed (dashed lines). }\label{fig:devband}
\end{figure}

\subsection{Gaussian wavepackets as charge carriers}
To simulate a flying-qubit implementation of the conditional phase shifter, we encode the electrons in Gaussian wavepackets of edge states belonging to the same Landau level $n$:
\begin{equation}
\psi_{\alpha}(x,y)=\int dk F(k,k^0_{\alpha},\sigma_{\alpha})\varphi_{n,k}(x)e^{iky}.\label{eq:eswp}
\end{equation}
The weight function $F(k,k^0_{\alpha},\sigma_{\alpha})=C\exp(-\sigma_{\alpha}^2(k-k^0_{\alpha})^2)$ linearly combines edge states with different wave vectors $k$ and a given Landau index $n$. 
The index $\alpha$ labels the translationally invariant region where the single-electron wavepacket is initialized.
The smoothness of the confining edge barrier, $V_{\alpha}(x)$, together with the central wavevector, $k^0_{\alpha}$ and the real-space broadening $\sigma_{\alpha}$ of the wavepacket, determine the group velocity $v_g^{\alpha}$ of the electron state. 
As proved in the numerical simulations of single and two-electron dynamics in Refs.\citep{BELLENTANIPRB2019, BELLENTANIPRB2018, BEGGIJOPCM2015}, the Gaussian wavepackets in Eq.~(\ref{eq:eswp}) maintain all the properties of edge states, i.e. the chirality and immunity to backscattering.
Moreover, the Gaussian shape of the single-electron state is preserved much more efficiently with respect to alternative frameworks, as in the presence of a  Lorentzian or exponential distribution in the energies. 

In the present geometry, we identify two translationally-invariant regions, $V_{\alpha}$ and $V_{\beta}$, where the single-electron wavepacket is initialized. The two interacting electrons are assumed to be injected in counterpropagating channels, and therefore the confining potentials must be characterized by the same smoothness but opposite bending, as in the geometry of the HOM interferometer from Ref.\cite{BELLENTANIPRB2019}. 
In order to include the fermionic antisimmetry of the two-electron wave function, the orbital single-electron wavepackets $\psi_{\alpha}$ and $\psi_{\beta}$ generated in the two initialization regions of the device are combined in the antisymmetric form
\begin{equation}
\Psi({\bf r_1},{\bf r_2})=\frac{\psi_{\alpha}({\bf r_1})\psi_{\beta}({\bf r_2})-\psi_{\beta}({\bf r_1})\psi_\alpha({\bf r_2})}{\sqrt{2}}.\label{eq:twoestate}
\end{equation}
We stress that the above wave function depends on four real-space coordinates: the memory burden needed to allocate numerically the corresponding 4D array (about $1$~Terabyte in our numerical simulations) can be afforded only by exploiting the resources of supercomputing facilities with memory-distributed architectures, and parallel techniques for high-performance computing. In particular, we distribute the two-particle wave function on a Cartesian topology of MPI processes, which maps the domain of the second particle $(x_2,y_2)$. 

\subsection{The Split-Step method for time evolution}
In our dynamic approach, the two-electron wave function $\Psi({\bf r_1},{\bf r_2})$ at initial time $t=0$ in Eq.~(\ref{eq:twoestate}) is evolved by iteratively applying the evolution operator
\begin{equation}
\hat{U}(\delta t)=e^{-i\frac{\hat{H}_{12}\cdot \delta t}{\hbar}},
\end{equation}
with the Hamiltonian $\hat{H}_{12}$ defined in Eq.~(\ref{eq:H2p}). 
In particular, we adapt the Split-Step Fourier method and the Trotter-Suzuky factorization to the present case of two interacting charges, as detailed in the following. 
 
In presence of non-negligible electron-electron interaction, $\hat{H}_{12}(x_1,y_1,x_2,y_2)$ can be rewritten as
\begin{equation}
\hat{H}_{12}=\hat{V}({\bf r}_1)+\hat{V}({\bf r}_2)+\hat{V}_{12}({\bf r}_1,{\bf r}_2)+\hat{T}_{x}(x_1,x_2)+\hat{T}_{y}(y_1,y_2),
\end{equation} 
with $\hat{V}$ single-particle external potential and 
\begin{eqnarray}
\hat{T}_{x}&=&\frac{\hat{p}_{x_1}^2}{2m^*}+\frac{\hat{p}_{x_2}^2}{2m^*}, \label{eq:kinx}\\
\hat{T}_{y}&=&\frac{\hat{p}_{y_1}^2}{2m^*}+\frac{2eB\hat{x}_{1}\hat{p}_{y_1}}{2m^*}+\frac{e^2B^2\hat{x}_1^2}{2m^*} \nonumber \\
&+&\frac{\hat{p}_{y_2}^2}{2m^*}+\frac{2eB\hat{x}_{2}\hat{p}_{y_2}}{2m^*}+\frac{e^2B^2\hat{x}_2^2}{2m^*}.\label{eq:kiny}
\end{eqnarray}
Equations~(\ref{eq:kinx}) and (\ref{eq:kiny}) represent the 2D kinetic operators for two free electrons in a perpendicular magnetic field, projected on the $\hat{x}$- or $\hat{y}$-direction of the real space. 
The potential operator $\hat{V}$ is characterized by a diagonal representation in the real space, $x_1y_1$ and $x_2y_2$.
The kinetic operator $\hat{T}_{x_1,x_2}$ is represented by a diagonal matrix in the 2D reciprocal space $k_{x_1}k_{x_2}$. The operator $\hat{T}_{y_1,y_2}$ is diagonal only in the 4D reciprocal space defined by $x_1k_{y_1}x_2k_{y_2}$. Finally, $\hat{V}_{12}$ couples the $\hat{x}$ and $\hat{y}$ coordinates, so that its diagonal representation is possible only in the 4D configuration space $x_1y_1x_2y_2$, which is the domain of the two-particle wavefunction.

According to the Trotter-Suzuky factorization method\citep{Kramer2010_PS}, the $\hat{U}(t;0)$ operator for an evolution time $t=N\cdot\delta t$ is factorized into three exponentials:\begin{equation}
[e^{-\frac{i}{\hbar}\delta t \hat{H}_{12}}]^N=[e^{-\frac{i}{\hbar}\delta t \cdot (\hat{V}_1+\hat{V}_2+\hat{V}_{12})}e^{-\frac{i}{\hbar}\delta t \cdot \hat{T}_{x}}e^{-\frac{i}{\hbar}\delta t \cdot \hat{T}_{y}}]^N,
\end{equation}
so that Fourier transforms $\textsf{F}_{x_1,x_2}(\textsf{F}_{y_1,y_2})$ and antitrasforms $\textsf{F}^{-1}_{x_1,x_2}(\textsf{F}^{-1}_{y_1,y_2})$ can be applied to switch from the real to the reciprocal space and to exploit the locality of the modified kinetic operators $\hat{T}_{x}(x_1,x_2)$ ($\hat{T}_{y}(y_1,y_2)$) in the reciprocal space $[k_{x_1},k_{x_2}]$ ($[k_{y_1},k_{y_2}]$). 
The evolution operator finally reads
\begin{align}\label{eq:evoloperat}
\hat{U}(t,0)=&[e^{-\frac{i}{\hbar}\delta t\cdot (\hat{V}_1+\hat{V}_2+\hat{V}_{12})}\textsf{F}^{-1}_{y_1,y_2}e^{-\frac{i}{\hbar}\delta t\cdot\hat{T}_{y_1,y_2}} \nonumber \\
&\textsf{F}_{y_1,y_2}\textsf{F}_{x_1,x_2}^{-1}e^{-\frac{i}{\hbar}\delta t\cdot \hat{T}_{x_1,x_2}}\textsf{F}_{x_1,x_2}]^N.
\end{align}

\begin{figure*}[t]
\centering
\includegraphics[width=1.\textwidth]{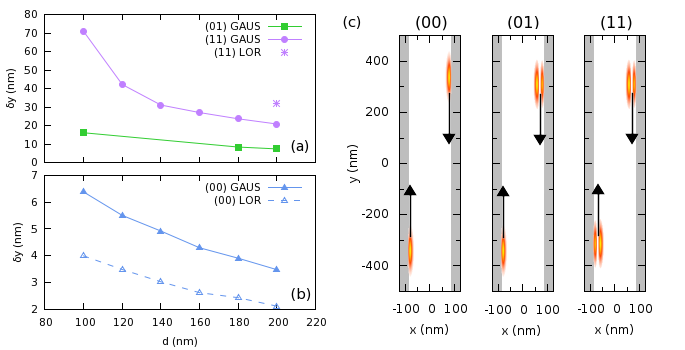}
\caption{(a) Spatial shift $\delta y$ for two indistinguishable electrons in the two excited configurations (01) and (11), with $\sigma = 40$~nm and in the presence of the long-range unscreened Coulomb interaction of Eq.~(\ref{eq:coulun}) with $d_z=1$~nm.  The label \textit{GAUS} refers to a Gaussian single-electron wavefunction, while \textit{LOR} to a lorentzian wavepacket.  (b) Spatial shift $\delta y$ for two indistinguishable electrons in the ground configuration (00) with the same parameters as above.  (c) Map of the active region of the device with the external potential (grey) and initial density probability distributions of two single-electron wavepackets (red) in the three configurations under study: (00) both electrons in the ground channel 0, (01) one electron in channel 0 and one electron in the excited channel 1, (11) both electrons in channel 1.  Black arrows define the direction of propagation of the wavepackets.}\label{fig:dyall}
\end{figure*}

\subsection{Numerical modeling of the active region}
We initially describe the active region of the device in Fig.~\ref{fig:realdev}(a) with a simplified geometry to expose the selectivity of Coulomb coupling for two electrons initialized in the first two Landau levels $n=0,1$.
In this model, the two-electron wave function propagate in a confined 2DEG with two sharp barriers in the transverse direction of the device. Thus, the external potential in the single-particle hamiltonian is $V(x)=V_b\left[\Theta(x_L-x)+\Theta(x-x_R)\right]$, where $V_b$ is the height of the barrier, $x_L$ and $x_R$ identify the turning points at the edges, and $\Theta(x)$ is the Heaviside function. 

Figure~\ref{fig:devband} compares the bandstructure of the first three Landau levels $E_{0}(x_0)$, $E_{1}(x_0)$ and $E_{2}(x_0)$ (solid lines) with the transverse shape of the external potential profile $\hat{V}(x)$ at the right side of the initialization region. 
The dots on the two Landau levels identify the centers $x_0(k)$ of the eigenstates involved in the linear combination of Eq.~(\ref{eq:eswp}) for two single-electron wavepackets in the ground ($n=0$) or in the excited ($n=1$) channel. 
The present device operates at bulk filling factor 2, i.e. the first two Landau levels are available at the energies involved in our operating regime, which are well below the third Landau level with a minimum energy $E_2 = 21$~meV. 
We observe that, for a central energy $E^0\approx 15$~meV and a real-space broadening $\sigma=40$~nm, the first Landau level is characterized by a sharper bending, thus resulting in a higher group velocity (or, equivalently, a smaller magnetic mass) for a Gaussian wavepacket initialized in n=0 with respect to the same in n=1. 
We therefore expect that Coulomb repulsion determines a larger spatial shift $\delta y$ for two electrons localized in $n_1=n_2=1$ with respect to alternative configurations. 
This effect is further enhanced by the different probability distribution of the single-electron wavepackets with $n=0,1$ in the transverse direction of the device, displayed in Fig.~\ref{fig:devband}. 
The initial wavepacket with $n=1$ has a larger probability in the bulk of the 2DEG. 
Note, however, that both wavepackets are mostly localized at positive values of the real-space domain in the $\hat{x}$-direction, thus ensuring a negligible overlap with the counterpropagating state - which is symmetric with respect to the origin. 
Then, for a distance $d>100$~nm between the edges, the two-electron scattering is purely driven by Coulomb interaction, and no interchannel tunneling is present.

\section{RESULTS}\label{sec:res}
In the following, we discuss the selective coupling of two electrons in counterpropagating Gaussian wavepackets of edge state, by comparing 3 different configurations labelled as $(n_1,n_2)$ with $n_1=0,1$ and $n_2=0,1$, and simulate their Coulomb-driven scattering by assuming an unscreened long-range soft Coulomb interaction:
\begin{equation}
V_{12}=\frac{e^2}{4\pi\epsilon_r\sqrt{(x_1-x_2)^2+(y_1-y_2)^2+d_z^2}},\label{eq:coulun}
\end{equation}
where $d_z=1$~nm avoids the divergence at ${\bf r}_1 = {\bf r}_2$ with a negligible effect on the numerical results, and $\epsilon_r$ is the medium permittivity of GaAs.

\subsection{Exact two-electron scattering in 2D}
Due to the exchange symmetry, we expect that at the final time (when $V_c$ is negligible due to the large distance between the two electrons) the probability of one of the two particles integrated over the other one shows the same value in the two outputs, as in the initial condition.
The effect of Coulomb interaction is then traced to the difference between the final density probability of the second particle in the interacting case and in the non-interacting case, in the direction of propagation (i.e. the $\hat{y}$-direction of the device).  
Indeed, when two electrons approach each other, electron-electron repulsion transforms part of their kinetic energy into inter-particle potential energy, so that the velocity along their path is reduced. 
As the relative distance returns then to the original value, the potential energy is transformed back into kinetic energy and the initial velocity is restored. 
This turns into a delay in the propagation of the two electrons compared to the non-interacting case, that corresponds to a phase factor in front of the $|11\rangle$ component of the two-qubit wavefunction in the global $\hat{T}$ transformation.

The difference between the maximum of this distribution in the interacting and non-interacting scenario, namely $\delta y$, depends on the geometry of the active region. 
The latter can be strongly affected by the distance between the two lateral edges, the wavepacket size and the smoothness of the confining barriers. 
With regards to the last parameter, a proper design of $V(x)$ with sharp lateral barriers is necessary to induce a quasi-parabolic dispersion of the second or first Landau level, rather than a linear one, to ensure a measurable $\delta y$. 
Indeed, in the presence of linear dispersion, the change in the kinetic energy of the two wavepackets does not change the velocity, so that the displacement $\delta y$ is zero.

Figures~\ref{fig:dyall}(a)-(b) compare the values of $\delta y$ in the 3 configurations, namely (00), (01) and (11), that are displayed in Fig.~\ref{fig:dyall}(c) for a range of distances between the edges, ranging from $d=100$ to $d=200$~nm. 
Note that, as described above, no tunneling is present in this operating regime. 
The effect of Coulomb repulsion generally decreases with the distance between the two borders of the device, $d$, and determines a larger longitudinal shift $\delta y$ for the configuration with two electrons in the excited channel $n = 1$. 
This trend agrees with the lower group velocity in the excited edge channel, together with a larger Coulomb interaction due to the transverse spatial distribution of the wavepacket for $n=1$ in Fig.~\ref{fig:devband}. 
The discrepancy between the values of $\delta y$ in the three cases increases significantly when the distance between the two edges is reduced.

\begin{figure}
\begin{minipage}{1.\columnwidth}
\includegraphics[width=0.95\columnwidth]{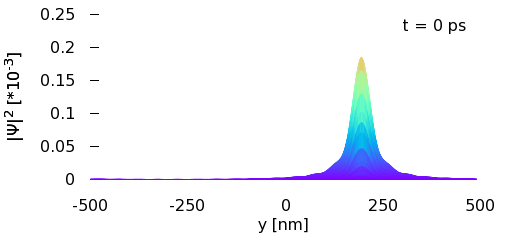}
\end{minipage}
\begin{minipage}{1.\columnwidth}
\includegraphics[width=0.95\columnwidth]{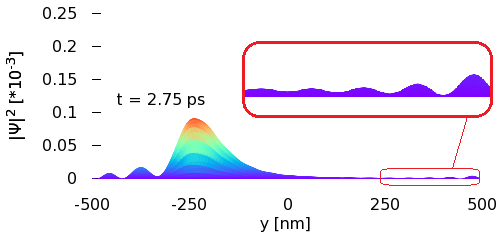}
\end{minipage}
\caption{Single-electron density probability for a Lorentzian wavepacket with $\Gamma=50$~nm and $n=0$ at (a) initial and (b) final time.}\label{fig:lorentz}
\end{figure}

We also remark that the functioning of this solid-state implementation of the conditional phase shifter does not depend qualitatively on the shape of the wavepacket. 
To prove this, we simulate the interaction of two Lorentzian single-electron wavepackets in our device, by adopting the numerical model validated for levitons\cite{DUBOISNATURE2013,RonettiPRR2020} in Ref.~\cite{KEELINGPRL2006}. 
In this scenario, the weight function $F(k)$ has an exponential distribution on the Fourier space, so that the wave function is Lorentzian in the real-space. 
Note that the present single-electron excitation operates in the same regime of the Gaussian wavepackets. 
Figure~\ref{fig:lorentz}(a) displays the initial distribution of the density probability in the longitudinal direction of the real space, while panel (b) shows the same density probability after its evolution in time. 
Differently from the Gaussian excitation, the Lorentzian pulse does not maintain its shape during the evolution, and oscillations are present in the tail of the density probability. 
Figure~\ref{fig:dyall} also displays the values of the spatial shift generated by a two-electron scattering in the active region for a Lorentzian wavepacket. 
By simulating the long-range Coulomb-driven interaction for two wavepackets in the $(00)$ configuration (Fig.~\ref{fig:dyall}(b)), we observe that the trend of the spatial shift do not differ qualitatively from the one computed with the Gaussian modelling of the electron state. 
Figure~\ref{fig:dyall}(a) further shows the value of $dy$ computed by studying the interaction of two Lorentzian states initialized in the excitated channel ($n=1$). 
In the present case, the computed value for the phase shift is larger than the one predicted in the Gaussian case; this follows from the lower group velocity of the Lorentzian excitation, whose dynamics is more strongly affected by the non linerarity of the bandstructure due to the large number of edge states in the tails of the weight distribution. 
However, also in the leviton-like modeling of single-electron wavepackets, the values of the spatial shifts in the ground and in the excited channel differ by tens of nanometers. 

\begin{figure}[t]
\centering
\includegraphics[width=1.\columnwidth]{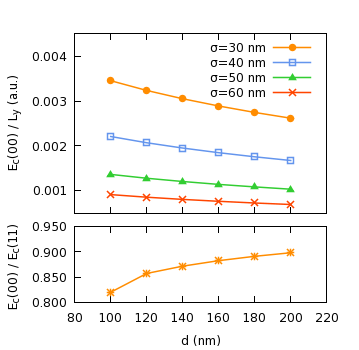}
\caption{(Top) integrated Coulomb energy per length size $\mathcal{E}_c$ in Eq.~(\ref{eq:intcoulen}) exchanged during the scattering of two indistinguishable electrons in the (00) configuration as a function of the distance between the edges of the confining potential $d$ for different wavepacket sizes $\sigma$. (Bottom) Ratio between the integrated Coulomb energy per length size $\mathcal{E}_c$ in the (00) configuration and in the (11) configuration.}\label{fig:energ}
\end{figure}

\subsection{Estimate of the interaction energy}
To analyze the interplay between the effects related to the different bandstructures of the Landau levels and the real-space distribution of the charge encoded in the Gaussian wavepackets, we estimate the total amount of energy exchanged during the two-electron scattering in the active region. 
We initially map the Coulomb potential $V_{12}(x_1,y_1,x_2,y_2)$ by fixing the $x_1$ and $x_2$ coordinates to the maxima of the single-electron wavepackets, namely $x_1=x_1^M$ in the channel on the right and $x_2=x_2^M$ in the channel on the left (see Fig.\ref{fig:dyall}(c)).
The Coulomb potential energy $V(x_1^M, y_1, x_2^M, y_2)$ is then averaged over the discrete set of $\hat{y}-$coordinates that define the path of each edge channels, i.e. $y_1=-y_2=Y$ with $Y \in [-L_y,L_y]$, where $L_y$ is the positive coordinate of the boundary in the $\hat{y}-$direction. 

To provide a better estimate of the energy exchanged, we also account for the 2D spatial distribution of the single charge. 
Therefore, we average the Coulomb potential $V(x_1^*, Y; x_2^*, -Y)$, on a 2D Gaussian distribution $F(x,x^*,y,y^*)$, that is centered in $x^*=x^M$ and $y^*=Y$, i.e.
\begin{equation}
    F(x,x^M,y,Y)=a_xe^{-(x-x^M)^2/2\sigma_x^2}a_ye^{-(y-Y)^2/2\sigma_y^2},\label{eq:2DGauss}
\end{equation}
where $\sigma_x$ and $\sigma_y$ are the real-space broadenings in the transverse and longitudinal direction respectively, and $a_x$ and $a_y$ are normalization constants.
The integrated Coulomb energy exchanged during the scattering per length size then reads:
\begin{eqnarray}
    \mathcal{E}_c=&&\frac{1}{L_y}\int dY d{\bf r}_1d{\bf r}_2 F(x_1,x_1^M,y_1,Y) \nonumber \\ &&V(x_1^M, Y, x_2^M, -Y)F(x_2,x_2^M,y_2,-Y).\label{eq:intcoulen}
\end{eqnarray}
$\mathcal{E}_c$ is displayed in the top panel of Fig.~\ref{fig:energ} as a function of the distance $d$ for different values of $\sigma$. 
Here, we approximate the spatial distribution of a single-electron wavepacket with $n=0$ by means of the above Gaussian distribution $F(x,x^M,y,Y)$ with $\sigma_x=5$~nm and $\sigma_y=\sigma$. 

The integrated Coulomb energy $\mathcal{E}_c$ shows that for smaller wavepackets the repulsion is larger, as a result of the increase in charge localization. 
This difference decreases with the distance $d$, in agreement with the fact that Coulomb repulsion becomes less effective when the electrons are further separated. 
Note, however, that this simple model does not take into account the real shape of the edge states and their corresponding bandstructure; indeed, when the distance between the two edges is reduced, the minimum of the Landau level is raised, so that for a given distance of the center $x_0(k)$ to the turning point of the barrier, $x_L$ or $x_R$, the energy broadening of the wavepackets shifts to lower values, thus decreasing the group velocity of the wavepacket. 
This is expected to affect more strongly smaller wavepackets in the real space, and therefore at shorter distances $\delta y$ eventually increases with $\sigma$. 

Finally, the bottom panel of Fig.~\ref{fig:energ} displays the ratio between the integrated energy exchanged during the scattering in the (00) configuration and in the (11) configuration, which is - in our operating regime - between $80\%$ and $90\%$. 
The large discrepancy between the values of $\delta y$ in the two configurations does not find an explanation in the amount of Coulomb energy exchanged alone: the different bending of the Landau levels - and therefore the smaller group velocity of the electron in the excited channel - is the origin of this effect.
\begin{figure}[b]
\centering
\includegraphics[width=1.\columnwidth]{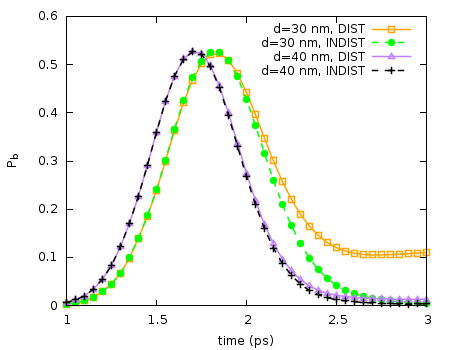}
\caption{Bunching probability Pb(t) during the Coulomb-driven scattering for two distinguishable/indistinguishable electrons initialized in Gaussian wavepackets with $\sigma = 40$~nm in the (00) configuration for small values of the distance $d$.  Note that tunneling between the two counterpropagating channel is present in this regime.}\label{fig:bunpc}
\end{figure}

\subsection{Screening and bunching probability}
Besides exposing the spatial shift generated by Coulomb interaction, the exact computation of the four-degrees-of-freedom wavefunction allows us to measure dynamically the two-electron bunching probability. The real-space domain is partitioned in a \textit{TOP} ($y>0$) and a \textit{BOTTOM} ($y<0$) region, which correspond to the outputs of the single-electron wavepackets $\psi_{\alpha}$ and $\psi_{\beta}$, respectively.  As in the electron HOM experiment in Ref.\cite{BELLENTANIPRB2019}, we compute the bunching probability as
\begin{equation}
    P_b=\int_{S_{TT}}dr_1dr_2|\Psi(r_1,r_2)|^2+\int_{S_{BB}}dr_1dr_2|\Psi(r_1,r_2)|^2,
\end{equation}
where $S_{TT}$ and $S_{BB}$ are 4D domains in the configuration space with $y_1,y_2>0$ and $y_1,y_2<0$, respectively. 
We estimate $P_b(t)$ in the two different cases of distinguishable and indistinguishable particles, for two small values of the distance $d$, namely $d=30$ and $d=40$~nm, and in the $(00)$ configuration. 
Note that, differently from the operating regime of the numerical simulations above (with $d>100$~nm), the two single-electron wavefunction in counterpropagating channels partially overlap in the transverse direction at initial time, so that tunneling between them could occurs during the scattering. 

\begin{figure}[t]
\centering
\includegraphics[width=1.\columnwidth]{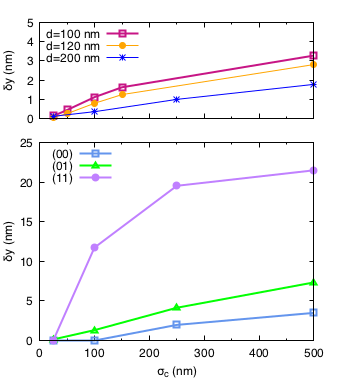}
\caption{(Top) Spatial shift $\delta y$ in presence of screening for two indistinguishable electrons initialized in wavepackets with $\sigma = 40$~nm in the (00) configuration. The value at $\sigma_c=500$~nm corresponds to $\delta y$ in presence of unscreened long-range interaction. (Bottom) Comparison between the spatial shifts $\delta y$ for two indistinguishable electrons with $\sigma = 40$~nm in the three different configurations and $d = 200$~nm with screening.}\label{fig:screening}
\end{figure}

The numerical results are displayed in Fig.~\ref{fig:bunpc}. 
By selecting a small value for $d_z$ in Eq.~(\ref{eq:coulun}), we ensure that the two electrons experience a very large Coulomb interaction at short distances. 
For the simulated values of the channel distance $d$ and distinguishable particles, we measure a non-zero bunching probability that generally decreases by reducing the distance, and it almost vanish for $d>40$~nm. 
We further observe that, as in the HOM geometry, Coulomb interaction does not fully reflect the two electron, differently from the expected result in an effective 1D geometry. 
In the latter scenario, indeed, the two electrons are forced to propagate on the same rail, so that at $y_1=y_2$ they experience diverging value of the Coulomb interaction. 
Moreover, with contrast to the case of Ref.~\cite{BELLENTANIPRB2019}, the bunching probability is fully quenched by the presence of exchange interaction, regardless the distance between the two rails. 
Differently from the wavepackets generated after the interaction with a quantum point contact, the reflected and transmitted states induced by Coulomb repulsion fully overlap in the Fourier space.

Finally, we simulate the effect of screening by adding an exponential damping in the Coulomb repulsion:
\begin{equation}
    V_{12}=\frac{Ce^2\exp(-\sqrt{(x_1-x_2)^2+(y_1-y_2)^2}/\sigma_c)}{4\pi\epsilon_r\sqrt{(x_1-x_2)^2+(y_1-y_2)^2+d_z^2}},\label{eq:vscreen}
\end{equation}
where $\sigma_c$ is the effective interaction length and $C$ the amplitude of the screening. 
Top panel of Fig.~\ref{fig:screening} shows the values of $\delta y$ in presence of screening for the (00) configuration with $\sigma=40$~nm, and a distance between the edges of the confining barrier that goes from $d=100$~nm to $d=200$~nm. 
In our operating regime, we simulate a damping length $\sigma_c$ of the order of the width of the active region ($L_x=250$~nm), and observe that the values of $\delta y$ are reduced by a factor -at least- $1/2$. 
The bottom panel of Fig.~\ref{fig:screening} compares the spatial shift $\delta y$ in the 3 configurations for the largest distance between the edges of the device ($d=200$~nm), that ensures the absence of interchannel tunneling in all configurations. 
We observe that, for the present values of $\sigma_C$, the (11) configuration is characterized by a spatial shift $\delta y$ that is large and visible in our numerical simulations. 

The above results predict that by properly tuning the geometrical parameters of the active region with modulation gates, or by varying the effective length for the screening, e.g. by modifying the electron density of the 2DEG, it is possible to quench the effect of Coulomb repulsion for all configurations except the (11) one, so that electron-electron repulsion acts as a selective entangler also for this simple geometry. The components of the latter device have dimentions that are feasible with current nanotechnology. 

In the next section we further simulate the Coulomb-driven scattering between two indistinguishable electrons with a more realistic geometry of the active region, that corresponds to the inner part of the loop area in the multichannel Mach-Zehnder interferometer of Ref.\cite{BELLENTANIPRB2018}; we then predict the phase shift $\gamma$ that rotates the (11) component of the two-electron wavefunction in the full-scale conditional phase shifter of Fig.~\ref{fig:realdev}(a).

\subsection{The phase shift $\gamma$ in the {\bf T} transformation}
We now address a more realistic profile of the confining potential and predict the phase $\gamma$ in the $T$ matrix of Eq.~(\ref{eq:tmatrix}) for the full-scale device of Fig.~\ref{fig:realdev}(a). 
First, we model the confining barrier with a smoothed profile in the $\hat{x}-$direction:
\begin{equation}
    V_{ext}(x)=V_b\left(\frac{1}{1+e^{\frac{x-x_{b}}{\lambda}}}+\frac{1}{1+e^{\frac{-x+x_{b}}{\lambda}}}\right),\label{eq:Vfermicphsh}
\end{equation}
where $V_b=0.31$~eV, $\lambda=3$~nm and $x_b=55$~nm. 
Figure~\ref{fig:realdev}(b) compares the potential profile in the transverse direction of the device (blue shaded area), to the bandstructure of the second Landau level (blue solid line) and the density probability of a single electron wavepacket with $n=1$, $\sigma=40$~nm, and an injection energy of $E^0=20.4$~meV (black dashed line). 
Note that this regime reproduces the geometry and the injection protocol of the multichannel MZI described in Ref~\cite{BELLENTANIPRB2018}, which is the building block of our proposal for a solid-state implementation of the conditional phase shifter.

Consistently to the findings in the simplified model presented above, we expect a stronger Coulomb interaction between the counterpropagating electrons with both Landau levels $n=1$. 
The sharper bending of the first Landau level determines a smaller magnetic mass $m^*_B$ with respect to a wavepacket with the same energy distribution but higher cyclotron index, $n$. 
The smaller group velocity for $n=1$ induces a larger shift in the real-space $\delta y$. 
Moreover, at a given value of the injection energy $E^0(k)$, the center $x_0^n(k)$ of an edge state with $n=0$ is closer to the profile of the confining barrier with respect to the corresponding edge state with $n=1$. 
The transverse probability distribution of two wavepackets in the second Landau level is then larger in the bulk with respect to the case of two wavepackets initialized in the ground state, thus enhancing the effect of Coulomb interaction.

The relation between the spatial shift $\delta y$ and the $\gamma$ factor in the ${\bf{T}}$ matrix can not be trivially determined by the wavevector $k$ alone, due to its gauge dependence. 
We resort to the difference in the optical paths $L_{eff}$, that is necessary to produce a $2\pi$ rotation in each single-electron MZI at the edge of the device. 

By means of a single-particle solver, we simulate the dynamics of single-electron interference in one of the two Mach-Zehnder interferometers reported in Fig.~\ref{fig:realdev}(a). 
Here, we artificially introduce a relative shift in the $\hat{y}-$direction ($\Delta Y$) between the two components of the wavefunction in the loop area. 
This spatial shift corresponds to a relative phase factor in the single-electron wavefunction between the transmitted wavepacket in $n=0$ and the transmitted one in $n=1$ after the scattering with the potential dip labeled as BS1 in Fig.\ref{fig:realdev}(a). 
In a full-scale two-qubit device, the introduction of an artificial shift $\Delta Y$ mimics the effect of Coulomb repulsion on the electron in the excited edge channel, when the counterpropagating electron is in the active region (yellow box in Fig\ref{fig:realdev}(a)). 
In the equivalent single-electron simulation, $\Delta Y$ rotates the final state at the output of the MZI. 
The spatial periodicity $L_{eff}$ of the interference pattern in the transmission amplitude for the 0 channel, $T(\Delta y)$, is then related to $\gamma$: if $L_{eff}$ corresponds to a $2\pi$ rotation in the output state of a single-electron Mach-Zehnder experiment, the same rotation, $\gamma=2\pi$, is obtained by introducing a selective Coulomb repulsion that shifts the final position of the two wavepackets in $n=1$ by a factor $\delta y = L_{eff}$.  
Within the present regime, we measure an effective length $L_{eff}=20$~nm. 
The dynamical simulation of the Coulomb-driven scattering of two indistinguishable electrons in the second edge channel provides the shift $\delta y$ for the present realistic geometry, which is measured to be $\delta y=11$~nm.  

We finally relate the shift $\delta y$ to the corresponding $\gamma$ in the $\hat{T}$ transformation in Eq.~(\ref{eq:tmatrix}) by using $L_{eff}$ as a reference and resosrting to the following equation:
\begin{equation}
\gamma=2\pi\frac{\delta y}{L_{eff}},\label{eq:cophshs}
\end{equation}
which provides $\gamma=\pi$ in the full-scale conditional phase shifter. 
This proves the feasibility of a selective phase shifter with a factor $\pi$ in our geometry, where the smoothed barriers are characterized by a relative distance $W=110$~nm at $B=5$~T. 
A proper increase of the distance between the outer edges of the two mesas at the nanometer scale is expected to induce a rotation that ranges from $\pi/2$ to $2\pi$, thus making this device a viable approach for conditional phase shifting driven by Coulomb interaction.

\section{Conclusions}\label{sec:conc}
We have shown that Coulomb interaction between two charge carriers moving in two counterpropagating edge channels can induce a consistent and controllable phase shift in one of the four configurations of possible Landau levels occupancy.
By encoding a qubit state into the Landau level index degree of freedom of one of the carrier, whith only two Landau levels being energetically accessible, the above phase only applies to the (11) state, thus creating a two-qubit conditional phase shifter $T$.
The quantom gate $T$ can be adopted, in turn, as the two-qubit entangling transformation of a universal set of quantum gates.

Our simulations address the numerically exact propagation of the two-particle wave function in a full-scale geometry of a device operating at bulk filling factor 2.
Thus, all the real-space effects of the electron-electron mutual interaction are accounted for, including the generation of quantum correlations between the longitudinal degrees of freedon (i.e. the particle positions along the edge channels) and the effect of the finite localization of the carriers.
Indeed, we found that the higher the spatial localization, the stronger the effect of Coulomb interaction. 
Also, a Lorentzian shape of the charge carrier wave function gives rise to the same entanglement effect as a Gaussian wavepacket, although the low-energy tail reduces in average the phase of the $T$ transformation.

Most important, we demonstrated that the conditional phase generated by the multi-edge state device can be as large as $\pi$ and its value can be controlled by the static confinement potential. 

\section*{ACKNOWLEDGEMENTS}
This work has been co-funded by the European Union’s Horizon 2020 Research and Innovation Programme through the FET Open project IQubits under Grant Agreement N. 829005. We also acknowledge CINECA for HPC computing resources and technical support under the ISCRA C initiatives QUPIDO (HP10CPRABZ) and CHINHEX (HP10CEMC7B). PB and AB are members of gnfm-INdAM. We thank Prof. Xavier Oriols for useful discussions.

\bibliography{Paperbib}

\begin{thebibliography}{40}%
\makeatletter
\providecommand \@ifxundefined [1]{%
 \@ifx{#1\undefined}
}%
\providecommand \@ifnum [1]{%
 \ifnum #1\expandafter \@firstoftwo
 \else \expandafter \@secondoftwo
 \fi
}%
\providecommand \@ifx [1]{%
 \ifx #1\expandafter \@firstoftwo
 \else \expandafter \@secondoftwo
 \fi
}%
\providecommand \natexlab [1]{#1}%
\providecommand \enquote  [1]{``#1''}%
\providecommand \bibnamefont  [1]{#1}%
\providecommand \bibfnamefont [1]{#1}%
\providecommand \citenamefont [1]{#1}%
\providecommand \href@noop [0]{\@secondoftwo}%
\providecommand \href [0]{\begingroup \@sanitize@url \@href}%
\providecommand \@href[1]{\@@startlink{#1}\@@href}%
\providecommand \@@href[1]{\endgroup#1\@@endlink}%
\providecommand \@sanitize@url [0]{\catcode `\\12\catcode `\$12\catcode
  `\&12\catcode `\#12\catcode `\^12\catcode `\_12\catcode `\%12\relax}%
\providecommand \@@startlink[1]{}%
\providecommand \@@endlink[0]{}%
\providecommand \url  [0]{\begingroup\@sanitize@url \@url }%
\providecommand \@url [1]{\endgroup\@href {#1}{\urlprefix }}%
\providecommand \urlprefix  [0]{URL }%
\providecommand \Eprint [0]{\href }%
\providecommand \doibase [0]{http://dx.doi.org/}%
\providecommand \selectlanguage [0]{\@gobble}%
\providecommand \bibinfo  [0]{\@secondoftwo}%
\providecommand \bibfield  [0]{\@secondoftwo}%
\providecommand \translation [1]{[#1]}%
\providecommand \BibitemOpen [0]{}%
\providecommand \bibitemStop [0]{}%
\providecommand \bibitemNoStop [0]{.\EOS\space}%
\providecommand \EOS [0]{\spacefactor3000\relax}%
\providecommand \BibitemShut  [1]{\csname bibitem#1\endcsname}%
\let\auto@bib@innerbib\@empty
\bibitem [{\citenamefont {Grenier}\ \emph {et~al.}(2011)\citenamefont
  {Grenier}, \citenamefont {Hervé}, \citenamefont {Féve},\ and\ \citenamefont
  {Degiovanni}}]{GrenierMPLB2011}%
  \BibitemOpen
  \bibfield  {author} {\bibinfo {author} {\bibfnamefont {C.}~\bibnamefont
  {Grenier}}, \bibinfo {author} {\bibfnamefont {R.}~\bibnamefont {Hervé}},
  \bibinfo {author} {\bibfnamefont {G.}~\bibnamefont {Féve}}, \ and\ \bibinfo
  {author} {\bibfnamefont {P.}~\bibnamefont {Degiovanni}},\ }\href {\doibase
  10.1142/S0217984911026772} {\bibfield  {journal} {\bibinfo  {journal} {Modern
  Physics Letters B}\ }\textbf {\bibinfo {volume} {25}},\ \bibinfo {pages}
  {1053} (\bibinfo {year} {2011})}\BibitemShut {NoStop}%
\bibitem [{\citenamefont {Roussel}\ \emph {et~al.}(2017)\citenamefont
  {Roussel}, \citenamefont {Cabart}, \citenamefont {Fève}, \citenamefont
  {Thibierge},\ and\ \citenamefont {Degiovanni}}]{RousselPSSB2017}%
  \BibitemOpen
  \bibfield  {author} {\bibinfo {author} {\bibfnamefont {B.}~\bibnamefont
  {Roussel}}, \bibinfo {author} {\bibfnamefont {C.}~\bibnamefont {Cabart}},
  \bibinfo {author} {\bibfnamefont {G.}~\bibnamefont {Fève}}, \bibinfo
  {author} {\bibfnamefont {E.}~\bibnamefont {Thibierge}}, \ and\ \bibinfo
  {author} {\bibfnamefont {P.}~\bibnamefont {Degiovanni}},\ }\href {\doibase
  10.1002/pssb.201600621} {\bibfield  {journal} {\bibinfo  {journal} {physica
  status solidi (b)}\ }\textbf {\bibinfo {volume} {254}},\ \bibinfo {pages}
  {1600621} (\bibinfo {year} {2017})}\BibitemShut {NoStop}%
\bibitem [{\citenamefont {Glattli}\ and\ \citenamefont
  {Roulleau}(2017)}]{GlattliPSSB2017}%
  \BibitemOpen
  \bibfield  {author} {\bibinfo {author} {\bibfnamefont {D.~C.}\ \bibnamefont
  {Glattli}}\ and\ \bibinfo {author} {\bibfnamefont {P.~S.}\ \bibnamefont
  {Roulleau}},\ }\href {\doibase 10.1002/pssb.201600650} {\bibfield  {journal}
  {\bibinfo  {journal} {physica status solidi (b)}\ }\textbf {\bibinfo {volume}
  {254}},\ \bibinfo {pages} {1600650} (\bibinfo {year} {2017})}\BibitemShut
  {NoStop}%
\bibitem [{\citenamefont {Locane}\ \emph {et~al.}(2019)\citenamefont {Locane},
  \citenamefont {Brouwer},\ and\ \citenamefont {Kashcheyevs}}]{LOCANENJP2019}%
  \BibitemOpen
  \bibfield  {author} {\bibinfo {author} {\bibfnamefont {E.}~\bibnamefont
  {Locane}}, \bibinfo {author} {\bibfnamefont {P.}~\bibnamefont {Brouwer}}, \
  and\ \bibinfo {author} {\bibfnamefont {V.}~\bibnamefont {Kashcheyevs}},\
  }\href {\doibase 10.1088/1367-2630/ab3fbb} {\bibfield  {journal} {\bibinfo
  {journal} {New Journal of Physics}\ }\textbf {\bibinfo {volume} {21}}
  (\bibinfo {year} {2019}),\ 10.1088/1367-2630/ab3fbb}\BibitemShut {NoStop}%
\bibitem [{\citenamefont {Rodriguez}(2019)}]{RodriguezPhD2019}%
  \BibitemOpen
  \bibfield  {author} {\bibinfo {author} {\bibfnamefont {R.~H.}\ \bibnamefont
  {Rodriguez}},\ }\emph {\bibinfo {title} {{Relaxation of quasiparticles
  injected above the Fermi sea of a Quantum Hall edge channel}}},\ \href
  {https://tel.archives-ouvertes.fr/tel-02501196} {\bibinfo {type} {Theses}},\
  \bibinfo  {school} {{Universit{\'e} Paris-Saclay}} (\bibinfo {year}
  {2019})\BibitemShut {NoStop}%
\bibitem [{\citenamefont {Ji}\ \emph {et~al.}(2003)\citenamefont {Ji},
  \citenamefont {Chung}, \citenamefont {Sprinzak}, \citenamefont {Heiblum},
  \citenamefont {Mahalu},\ and\ \citenamefont {Shtrikman}}]{JINATURE2003}%
  \BibitemOpen
  \bibfield  {author} {\bibinfo {author} {\bibfnamefont {Y.}~\bibnamefont
  {Ji}}, \bibinfo {author} {\bibfnamefont {Y.}~\bibnamefont {Chung}}, \bibinfo
  {author} {\bibfnamefont {D.}~\bibnamefont {Sprinzak}}, \bibinfo {author}
  {\bibfnamefont {M.}~\bibnamefont {Heiblum}}, \bibinfo {author} {\bibfnamefont
  {D.}~\bibnamefont {Mahalu}}, \ and\ \bibinfo {author} {\bibfnamefont
  {H.}~\bibnamefont {Shtrikman}},\ }\href {\doibase 10.1038/nature01503}
  {\bibfield  {journal} {\bibinfo  {journal} {Nature}\ }\textbf {\bibinfo
  {volume} {422}},\ \bibinfo {pages} {415} (\bibinfo {year}
  {2003})}\BibitemShut {NoStop}%
\bibitem [{\citenamefont {Neder}\ \emph {et~al.}(2006)\citenamefont {Neder},
  \citenamefont {Heiblum}, \citenamefont {Levinson}, \citenamefont {Mahalu},\
  and\ \citenamefont {Umansky}}]{NEDERPRL2006}%
  \BibitemOpen
  \bibfield  {author} {\bibinfo {author} {\bibfnamefont {I.}~\bibnamefont
  {Neder}}, \bibinfo {author} {\bibfnamefont {M.}~\bibnamefont {Heiblum}},
  \bibinfo {author} {\bibfnamefont {Y.}~\bibnamefont {Levinson}}, \bibinfo
  {author} {\bibfnamefont {D.}~\bibnamefont {Mahalu}}, \ and\ \bibinfo {author}
  {\bibfnamefont {V.}~\bibnamefont {Umansky}},\ }\href {\doibase
  10.1103/PhysRevLett.96.016804} {\bibfield  {journal} {\bibinfo  {journal}
  {Phys. Rev. Lett.}\ }\textbf {\bibinfo {volume} {96}},\ \bibinfo {pages}
  {016804} (\bibinfo {year} {2006})}\BibitemShut {NoStop}%
\bibitem [{\citenamefont {Kang}(2007)}]{KANGPRB2007}%
  \BibitemOpen
  \bibfield  {author} {\bibinfo {author} {\bibfnamefont {K.}~\bibnamefont
  {Kang}},\ }\href {\doibase 10.1103/PhysRevB.75.125326} {\bibfield  {journal}
  {\bibinfo  {journal} {Phys. Rev. B}\ }\textbf {\bibinfo {volume} {75}},\
  \bibinfo {pages} {125326} (\bibinfo {year} {2007})}\BibitemShut {NoStop}%
\bibitem [{\citenamefont {Lepage}\ \emph {et~al.}(2020)\citenamefont {Lepage},
  \citenamefont {Lasek}, \citenamefont {Arvidsson-Shukur},\ and\ \citenamefont
  {Barnes}}]{LepagePRA2020}%
  \BibitemOpen
  \bibfield  {author} {\bibinfo {author} {\bibfnamefont {H.~V.}\ \bibnamefont
  {Lepage}}, \bibinfo {author} {\bibfnamefont {A.~A.}\ \bibnamefont {Lasek}},
  \bibinfo {author} {\bibfnamefont {D.~R.~M.}\ \bibnamefont
  {Arvidsson-Shukur}}, \ and\ \bibinfo {author} {\bibfnamefont {C.~H.~W.}\
  \bibnamefont {Barnes}},\ }\href {\doibase 10.1103/PhysRevA.101.022329}
  {\bibfield  {journal} {\bibinfo  {journal} {Phys. Rev. A}\ }\textbf {\bibinfo
  {volume} {101}},\ \bibinfo {pages} {022329} (\bibinfo {year}
  {2020})}\BibitemShut {NoStop}%
\bibitem [{\citenamefont {Roulleau}\ \emph {et~al.}(2008)\citenamefont
  {Roulleau}, \citenamefont {Portier}, \citenamefont {Roche}, \citenamefont
  {Cavanna}, \citenamefont {Faini}, \citenamefont {Gennser},\ and\
  \citenamefont {Mailly}}]{ROULLEAUPRL2008}%
  \BibitemOpen
  \bibfield  {author} {\bibinfo {author} {\bibfnamefont {P.}~\bibnamefont
  {Roulleau}}, \bibinfo {author} {\bibfnamefont {F.}~\bibnamefont {Portier}},
  \bibinfo {author} {\bibfnamefont {P.}~\bibnamefont {Roche}}, \bibinfo
  {author} {\bibfnamefont {A.}~\bibnamefont {Cavanna}}, \bibinfo {author}
  {\bibfnamefont {G.}~\bibnamefont {Faini}}, \bibinfo {author} {\bibfnamefont
  {U.}~\bibnamefont {Gennser}}, \ and\ \bibinfo {author} {\bibfnamefont
  {D.}~\bibnamefont {Mailly}},\ }\href {\doibase
  10.1103/PhysRevLett.100.126802} {\bibfield  {journal} {\bibinfo  {journal}
  {Phys. Rev. Lett.}\ }\textbf {\bibinfo {volume} {100}},\ \bibinfo {pages}
  {126802} (\bibinfo {year} {2008})}\BibitemShut {NoStop}%
\bibitem [{\citenamefont {Deviatov}\ \emph {et~al.}(2011)\citenamefont
  {Deviatov}, \citenamefont {Ganczarczyk}, \citenamefont {Lorke}, \citenamefont
  {Biasiol},\ and\ \citenamefont {Sorba}}]{DEVIATOPRB2011}%
  \BibitemOpen
  \bibfield  {author} {\bibinfo {author} {\bibfnamefont {E.~V.}\ \bibnamefont
  {Deviatov}}, \bibinfo {author} {\bibfnamefont {A.}~\bibnamefont
  {Ganczarczyk}}, \bibinfo {author} {\bibfnamefont {A.}~\bibnamefont {Lorke}},
  \bibinfo {author} {\bibfnamefont {G.}~\bibnamefont {Biasiol}}, \ and\
  \bibinfo {author} {\bibfnamefont {L.}~\bibnamefont {Sorba}},\ }\href
  {\doibase 10.1103/PhysRevB.84.235313} {\bibfield  {journal} {\bibinfo
  {journal} {Phys. Rev. B}\ }\textbf {\bibinfo {volume} {84}},\ \bibinfo
  {pages} {235313} (\bibinfo {year} {2011})}\BibitemShut {NoStop}%
\bibitem [{\citenamefont {Beggi}\ \emph {et~al.}(2015)\citenamefont {Beggi},
  \citenamefont {Bordone}, \citenamefont {Buscemi},\ and\ \citenamefont
  {Bertoni}}]{BEGGIJOPCM2015}%
  \BibitemOpen
  \bibfield  {author} {\bibinfo {author} {\bibfnamefont {A.}~\bibnamefont
  {Beggi}}, \bibinfo {author} {\bibfnamefont {P.}~\bibnamefont {Bordone}},
  \bibinfo {author} {\bibfnamefont {F.}~\bibnamefont {Buscemi}}, \ and\
  \bibinfo {author} {\bibfnamefont {A.}~\bibnamefont {Bertoni}},\ }\href
  {\doibase 10.1088/0953-8984/27/47/475301} {\bibfield  {journal} {\bibinfo
  {journal} {Journal of Physics: Condensed Matter}\ }\textbf {\bibinfo {volume}
  {27}},\ \bibinfo {pages} {475301} (\bibinfo {year} {2015})}\BibitemShut
  {NoStop}%
\bibitem [{\citenamefont {Giovannetti}\ \emph {et~al.}(2008)\citenamefont
  {Giovannetti}, \citenamefont {Taddei}, \citenamefont {Frustaglia},\ and\
  \citenamefont {Fazio}}]{GIOVANNETTIPRB2008}%
  \BibitemOpen
  \bibfield  {author} {\bibinfo {author} {\bibfnamefont {V.}~\bibnamefont
  {Giovannetti}}, \bibinfo {author} {\bibfnamefont {F.}~\bibnamefont {Taddei}},
  \bibinfo {author} {\bibfnamefont {D.}~\bibnamefont {Frustaglia}}, \ and\
  \bibinfo {author} {\bibfnamefont {R.}~\bibnamefont {Fazio}},\ }\href
  {\doibase 10.1103/PhysRevB.77.155320} {\bibfield  {journal} {\bibinfo
  {journal} {Phys. Rev. B}\ }\textbf {\bibinfo {volume} {77}},\ \bibinfo
  {pages} {155320} (\bibinfo {year} {2008})}\BibitemShut {NoStop}%
\bibitem [{\citenamefont {Karmakar}\ \emph {et~al.}(2015)\citenamefont
  {Karmakar}, \citenamefont {Venturelli}, \citenamefont {Chirolli},
  \citenamefont {Giovannetti}, \citenamefont {Fazio}, \citenamefont {Roddaro},
  \citenamefont {Pfeiffer}, \citenamefont {West}, \citenamefont {Taddei},\ and\
  \citenamefont {Pellegrini}}]{KARMAKARPRB2015}%
  \BibitemOpen
  \bibfield  {author} {\bibinfo {author} {\bibfnamefont {B.}~\bibnamefont
  {Karmakar}}, \bibinfo {author} {\bibfnamefont {D.}~\bibnamefont
  {Venturelli}}, \bibinfo {author} {\bibfnamefont {L.}~\bibnamefont
  {Chirolli}}, \bibinfo {author} {\bibfnamefont {V.}~\bibnamefont
  {Giovannetti}}, \bibinfo {author} {\bibfnamefont {R.}~\bibnamefont {Fazio}},
  \bibinfo {author} {\bibfnamefont {S.}~\bibnamefont {Roddaro}}, \bibinfo
  {author} {\bibfnamefont {L.~N.}\ \bibnamefont {Pfeiffer}}, \bibinfo {author}
  {\bibfnamefont {K.~W.}\ \bibnamefont {West}}, \bibinfo {author}
  {\bibfnamefont {F.}~\bibnamefont {Taddei}}, \ and\ \bibinfo {author}
  {\bibfnamefont {V.}~\bibnamefont {Pellegrini}},\ }\href {\doibase
  10.1103/PhysRevB.92.195303} {\bibfield  {journal} {\bibinfo  {journal} {Phys.
  Rev. B}\ }\textbf {\bibinfo {volume} {92}},\ \bibinfo {pages} {195303}
  (\bibinfo {year} {2015})}\BibitemShut {NoStop}%
\bibitem [{\citenamefont {Bellentani}\ \emph {et~al.}(2018)\citenamefont
  {Bellentani}, \citenamefont {Beggi}, \citenamefont {Bordone},\ and\
  \citenamefont {Bertoni}}]{BELLENTANIPRB2018}%
  \BibitemOpen
  \bibfield  {author} {\bibinfo {author} {\bibfnamefont {L.}~\bibnamefont
  {Bellentani}}, \bibinfo {author} {\bibfnamefont {A.}~\bibnamefont {Beggi}},
  \bibinfo {author} {\bibfnamefont {P.}~\bibnamefont {Bordone}}, \ and\
  \bibinfo {author} {\bibfnamefont {A.}~\bibnamefont {Bertoni}},\ }\href
  {\doibase 10.1103/PhysRevB.97.205419} {\bibfield  {journal} {\bibinfo
  {journal} {Phys. Rev. B}\ }\textbf {\bibinfo {volume} {97}},\ \bibinfo
  {pages} {205419} (\bibinfo {year} {2018})}\BibitemShut {NoStop}%
\bibitem [{\citenamefont {Oliver}\ \emph {et~al.}(1999)\citenamefont {Oliver},
  \citenamefont {Kim}, \citenamefont {Liu},\ and\ \citenamefont
  {Yamamoto}}]{OLIVERSCIENCE1999}%
  \BibitemOpen
  \bibfield  {author} {\bibinfo {author} {\bibfnamefont {W.~D.}\ \bibnamefont
  {Oliver}}, \bibinfo {author} {\bibfnamefont {J.}~\bibnamefont {Kim}},
  \bibinfo {author} {\bibfnamefont {R.~C.}\ \bibnamefont {Liu}}, \ and\
  \bibinfo {author} {\bibfnamefont {Y.}~\bibnamefont {Yamamoto}},\ }\href
  {\doibase 10.1126/science.284.5412.299} {\bibfield  {journal} {\bibinfo
  {journal} {Science}\ }\textbf {\bibinfo {volume} {284}},\ \bibinfo {pages}
  {299} (\bibinfo {year} {1999})}\BibitemShut {NoStop}%
\bibitem [{\citenamefont {B\"uttiker}\ \emph {et~al.}(2003)\citenamefont
  {B\"uttiker}, \citenamefont {Samuelsson},\ and\ \citenamefont
  {Sukhorukov}}]{BUTTIKERPHYSICAE2003}%
  \BibitemOpen
  \bibfield  {author} {\bibinfo {author} {\bibfnamefont {M.}~\bibnamefont
  {B\"uttiker}}, \bibinfo {author} {\bibfnamefont {P.}~\bibnamefont
  {Samuelsson}}, \ and\ \bibinfo {author} {\bibfnamefont {E.}~\bibnamefont
  {Sukhorukov}},\ }\href {\doibase https://doi.org/10.1016/j.physe.2003.09.019}
  {\bibfield  {journal} {\bibinfo  {journal} {Physica E: Low-dimensional
  Systems and Nanostructures}\ }\textbf {\bibinfo {volume} {20}},\ \bibinfo
  {pages} {33 } (\bibinfo {year} {2003})},\ \bibinfo {note} {proceedings of the
  International Symposium "Quantum Hall Effect: Past, Present and
  Future}\BibitemShut {NoStop}%
\bibitem [{\citenamefont {Chung}\ \emph {et~al.}(2005)\citenamefont {Chung},
  \citenamefont {Samuelsson},\ and\ \citenamefont {B\"uttiker}}]{CHUNGPRB2005}%
  \BibitemOpen
  \bibfield  {author} {\bibinfo {author} {\bibfnamefont {V.~S.-W.}\
  \bibnamefont {Chung}}, \bibinfo {author} {\bibfnamefont {P.}~\bibnamefont
  {Samuelsson}}, \ and\ \bibinfo {author} {\bibfnamefont {M.}~\bibnamefont
  {B\"uttiker}},\ }\href {\doibase 10.1103/PhysRevB.72.125320} {\bibfield
  {journal} {\bibinfo  {journal} {Phys. Rev. B}\ }\textbf {\bibinfo {volume}
  {72}},\ \bibinfo {pages} {125320} (\bibinfo {year} {2005})}\BibitemShut
  {NoStop}%
\bibitem [{\citenamefont {Samuelsson}\ \emph {et~al.}(2004)\citenamefont
  {Samuelsson}, \citenamefont {Sukhorukov},\ and\ \citenamefont
  {B\"uttiker}}]{SAMUELSSONPRL2004}%
  \BibitemOpen
  \bibfield  {author} {\bibinfo {author} {\bibfnamefont {P.}~\bibnamefont
  {Samuelsson}}, \bibinfo {author} {\bibfnamefont {E.~V.}\ \bibnamefont
  {Sukhorukov}}, \ and\ \bibinfo {author} {\bibfnamefont {M.}~\bibnamefont
  {B\"uttiker}},\ }\href {\doibase 10.1103/PhysRevLett.92.026805} {\bibfield
  {journal} {\bibinfo  {journal} {Phys. Rev. Lett.}\ }\textbf {\bibinfo
  {volume} {92}},\ \bibinfo {pages} {026805} (\bibinfo {year}
  {2004})}\BibitemShut {NoStop}%
\bibitem [{\citenamefont {Neder}\ \emph {et~al.}(2007)\citenamefont {Neder},
  \citenamefont {Ofek}, \citenamefont {Chung}, \citenamefont {Heiblum},
  \citenamefont {Mahalu},\ and\ \citenamefont {Umansky}}]{NEDERNATURE2007}%
  \BibitemOpen
  \bibfield  {author} {\bibinfo {author} {\bibfnamefont {I.}~\bibnamefont
  {Neder}}, \bibinfo {author} {\bibfnamefont {N.}~\bibnamefont {Ofek}},
  \bibinfo {author} {\bibfnamefont {Y.}~\bibnamefont {Chung}}, \bibinfo
  {author} {\bibfnamefont {M.}~\bibnamefont {Heiblum}}, \bibinfo {author}
  {\bibfnamefont {D.}~\bibnamefont {Mahalu}}, \ and\ \bibinfo {author}
  {\bibfnamefont {V.}~\bibnamefont {Umansky}},\ }\href {\doibase
  10.1038/nature05955} {\bibfield  {journal} {\bibinfo  {journal} {Nature}\
  }\textbf {\bibinfo {volume} {448}},\ \bibinfo {pages} {333} (\bibinfo {year}
  {2007})}\BibitemShut {NoStop}%
\bibitem [{\citenamefont {Glattli}\ and\ \citenamefont
  {Roulleau}(2016)}]{GLATTLISCIENCE2016}%
  \BibitemOpen
  \bibfield  {author} {\bibinfo {author} {\bibfnamefont {D.}~\bibnamefont
  {Glattli}}\ and\ \bibinfo {author} {\bibfnamefont {P.}~\bibnamefont
  {Roulleau}},\ }\href {\doibase https://doi.org/10.1016/j.physe.2015.10.034}
  {\bibfield  {journal} {\bibinfo  {journal} {Physica E: Low-dimensional
  Systems and Nanostructures}\ }\textbf {\bibinfo {volume} {76}},\ \bibinfo
  {pages} {216 } (\bibinfo {year} {2016})}\BibitemShut {NoStop}%
\bibitem [{\citenamefont {Bellentani}\ \emph {et~al.}(2019)\citenamefont
  {Bellentani}, \citenamefont {Bordone}, \citenamefont {Oriols},\ and\
  \citenamefont {Bertoni}}]{BELLENTANIPRB2019}%
  \BibitemOpen
  \bibfield  {author} {\bibinfo {author} {\bibfnamefont {L.}~\bibnamefont
  {Bellentani}}, \bibinfo {author} {\bibfnamefont {P.}~\bibnamefont {Bordone}},
  \bibinfo {author} {\bibfnamefont {X.}~\bibnamefont {Oriols}}, \ and\ \bibinfo
  {author} {\bibfnamefont {A.}~\bibnamefont {Bertoni}},\ }\href {\doibase
  10.1103/PhysRevB.99.245415} {\bibfield  {journal} {\bibinfo  {journal} {Phys.
  Rev. B}\ }\textbf {\bibinfo {volume} {99}},\ \bibinfo {pages} {245415}
  (\bibinfo {year} {2019})}\BibitemShut {NoStop}%
\bibitem [{\citenamefont {Ferraro}\ \emph {et~al.}(2018)\citenamefont
  {Ferraro}, \citenamefont {Ronetti}, \citenamefont {Vannucci}, \citenamefont
  {Acciai}, \citenamefont {Rech}, \citenamefont {Jockheere}, \citenamefont
  {Martin},\ and\ \citenamefont {Sassetti}}]{Ferraro2018_EuPJST}%
  \BibitemOpen
  \bibfield  {author} {\bibinfo {author} {\bibfnamefont {D.}~\bibnamefont
  {Ferraro}}, \bibinfo {author} {\bibfnamefont {F.}~\bibnamefont {Ronetti}},
  \bibinfo {author} {\bibfnamefont {L.}~\bibnamefont {Vannucci}}, \bibinfo
  {author} {\bibfnamefont {M.}~\bibnamefont {Acciai}}, \bibinfo {author}
  {\bibfnamefont {J.}~\bibnamefont {Rech}}, \bibinfo {author} {\bibfnamefont
  {T.}~\bibnamefont {Jockheere}}, \bibinfo {author} {\bibfnamefont
  {T.}~\bibnamefont {Martin}}, \ and\ \bibinfo {author} {\bibfnamefont
  {M.}~\bibnamefont {Sassetti}},\ }\href {\doibase
  10.1140/epjst/e2018-800074-1} {\bibfield  {journal} {\bibinfo  {journal} {The
  European Physical Journal Special Topics}\ }\textbf {\bibinfo {volume}
  {227}},\ \bibinfo {pages} {1345} (\bibinfo {year} {2018})}\BibitemShut
  {NoStop}%
\bibitem [{\citenamefont {Marguerite}\ \emph {et~al.}(2016)\citenamefont
  {Marguerite}, \citenamefont {Cabart}, \citenamefont {Wahl}, \citenamefont
  {Roussel}, \citenamefont {Freulon}, \citenamefont {Ferraro}, \citenamefont
  {Grenier}, \citenamefont {Berroir}, \citenamefont
  {Pla\ifmmode~\mbox{\c{c}}\else \c{c}\fi{}ais}, \citenamefont {Jonckheere},
  \citenamefont {Rech}, \citenamefont {Martin}, \citenamefont {Degiovanni},
  \citenamefont {Cavanna}, \citenamefont {Jin},\ and\ \citenamefont
  {F\`eve}}]{MARGUERITEPRB2016}%
  \BibitemOpen
  \bibfield  {author} {\bibinfo {author} {\bibfnamefont {A.}~\bibnamefont
  {Marguerite}}, \bibinfo {author} {\bibfnamefont {C.}~\bibnamefont {Cabart}},
  \bibinfo {author} {\bibfnamefont {C.}~\bibnamefont {Wahl}}, \bibinfo {author}
  {\bibfnamefont {B.}~\bibnamefont {Roussel}}, \bibinfo {author} {\bibfnamefont
  {V.}~\bibnamefont {Freulon}}, \bibinfo {author} {\bibfnamefont
  {D.}~\bibnamefont {Ferraro}}, \bibinfo {author} {\bibfnamefont
  {C.}~\bibnamefont {Grenier}}, \bibinfo {author} {\bibfnamefont {J.-M.}\
  \bibnamefont {Berroir}}, \bibinfo {author} {\bibfnamefont {B.}~\bibnamefont
  {Pla\ifmmode~\mbox{\c{c}}\else \c{c}\fi{}ais}}, \bibinfo {author}
  {\bibfnamefont {T.}~\bibnamefont {Jonckheere}}, \bibinfo {author}
  {\bibfnamefont {J.}~\bibnamefont {Rech}}, \bibinfo {author} {\bibfnamefont
  {T.}~\bibnamefont {Martin}}, \bibinfo {author} {\bibfnamefont
  {P.}~\bibnamefont {Degiovanni}}, \bibinfo {author} {\bibfnamefont
  {A.}~\bibnamefont {Cavanna}}, \bibinfo {author} {\bibfnamefont
  {Y.}~\bibnamefont {Jin}}, \ and\ \bibinfo {author} {\bibfnamefont
  {G.}~\bibnamefont {F\`eve}},\ }\href {\doibase 10.1103/PhysRevB.94.115311}
  {\bibfield  {journal} {\bibinfo  {journal} {Phys. Rev. B}\ }\textbf {\bibinfo
  {volume} {94}},\ \bibinfo {pages} {115311} (\bibinfo {year}
  {2016})}\BibitemShut {NoStop}%
\bibitem [{\citenamefont {Marian}\ \emph {et~al.}(2015)\citenamefont {Marian},
  \citenamefont {Colom{\'{e}}s},\ and\ \citenamefont
  {Oriols}}]{MARIANJPHYSCMAT2015}%
  \BibitemOpen
  \bibfield  {author} {\bibinfo {author} {\bibfnamefont {D.}~\bibnamefont
  {Marian}}, \bibinfo {author} {\bibfnamefont {E.}~\bibnamefont
  {Colom{\'{e}}s}}, \ and\ \bibinfo {author} {\bibfnamefont {X.}~\bibnamefont
  {Oriols}},\ }\href {\doibase 10.1088/0953-8984/27/24/245302} {\bibfield
  {journal} {\bibinfo  {journal} {Journal of Physics: Condensed Matter}\
  }\textbf {\bibinfo {volume} {27}},\ \bibinfo {pages} {245302} (\bibinfo
  {year} {2015})}\BibitemShut {NoStop}%
\bibitem [{\citenamefont {Wahl}\ \emph {et~al.}(2014)\citenamefont {Wahl},
  \citenamefont {Rech}, \citenamefont {Jonckheere},\ and\ \citenamefont
  {Martin}}]{WAHLPRL2014}%
  \BibitemOpen
  \bibfield  {author} {\bibinfo {author} {\bibfnamefont {C.}~\bibnamefont
  {Wahl}}, \bibinfo {author} {\bibfnamefont {J.}~\bibnamefont {Rech}}, \bibinfo
  {author} {\bibfnamefont {T.}~\bibnamefont {Jonckheere}}, \ and\ \bibinfo
  {author} {\bibfnamefont {T.}~\bibnamefont {Martin}},\ }\href {\doibase
  10.1103/PhysRevLett.112.046802} {\bibfield  {journal} {\bibinfo  {journal}
  {Phys. Rev. Lett.}\ }\textbf {\bibinfo {volume} {112}},\ \bibinfo {pages}
  {046802} (\bibinfo {year} {2014})}\BibitemShut {NoStop}%
\bibitem [{\citenamefont {Bertoni}\ \emph {et~al.}(2000)\citenamefont
  {Bertoni}, \citenamefont {Bordone}, \citenamefont {Brunetti}, \citenamefont
  {Jacoboni},\ and\ \citenamefont {Reggiani}}]{Bertoni_PRL2000}%
  \BibitemOpen
  \bibfield  {author} {\bibinfo {author} {\bibfnamefont {A.}~\bibnamefont
  {Bertoni}}, \bibinfo {author} {\bibfnamefont {P.}~\bibnamefont {Bordone}},
  \bibinfo {author} {\bibfnamefont {R.}~\bibnamefont {Brunetti}}, \bibinfo
  {author} {\bibfnamefont {C.}~\bibnamefont {Jacoboni}}, \ and\ \bibinfo
  {author} {\bibfnamefont {S.}~\bibnamefont {Reggiani}},\ }\href {\doibase
  10.1103/PhysRevLett.84.5912} {\bibfield  {journal} {\bibinfo  {journal}
  {Phys. Rev. Lett.}\ }\textbf {\bibinfo {volume} {84}},\ \bibinfo {pages}
  {5912} (\bibinfo {year} {2000})}\BibitemShut {NoStop}%
\bibitem [{\citenamefont {Bordone}\ \emph {et~al.}(2019)\citenamefont
  {Bordone}, \citenamefont {Bellentani},\ and\ \citenamefont
  {Bertoni}}]{BORDONESST2019}%
  \BibitemOpen
  \bibfield  {author} {\bibinfo {author} {\bibfnamefont {P.}~\bibnamefont
  {Bordone}}, \bibinfo {author} {\bibfnamefont {L.}~\bibnamefont {Bellentani}},
  \ and\ \bibinfo {author} {\bibfnamefont {A.}~\bibnamefont {Bertoni}},\ }\href
  {\doibase 10.1088/1361-6641/ab3be6} {\bibfield  {journal} {\bibinfo
  {journal} {Semiconductor Science and Technology}\ }\textbf {\bibinfo {volume}
  {34}} (\bibinfo {year} {2019}),\ 10.1088/1361-6641/ab3be6}\BibitemShut
  {NoStop}%
\bibitem [{\citenamefont {Chirolli}\ \emph {et~al.}(2013)\citenamefont
  {Chirolli}, \citenamefont {Taddei}, \citenamefont {Fazio},\ and\
  \citenamefont {Giovannetti}}]{CHIROLLIPRL2013}%
  \BibitemOpen
  \bibfield  {author} {\bibinfo {author} {\bibfnamefont {L.}~\bibnamefont
  {Chirolli}}, \bibinfo {author} {\bibfnamefont {F.}~\bibnamefont {Taddei}},
  \bibinfo {author} {\bibfnamefont {R.}~\bibnamefont {Fazio}}, \ and\ \bibinfo
  {author} {\bibfnamefont {V.}~\bibnamefont {Giovannetti}},\ }\href {\doibase
  10.1103/PhysRevLett.111.036801} {\bibfield  {journal} {\bibinfo  {journal}
  {Phys. Rev. Lett.}\ }\textbf {\bibinfo {volume} {111}},\ \bibinfo {pages}
  {036801} (\bibinfo {year} {2013})}\BibitemShut {NoStop}%
\bibitem [{\citenamefont {Kramer}\ \emph {et~al.}(2010)\citenamefont {Kramer},
  \citenamefont {Kreisbeck},\ and\ \citenamefont {Krueckl}}]{Kramer2010_PS}%
  \BibitemOpen
  \bibfield  {author} {\bibinfo {author} {\bibfnamefont {T.}~\bibnamefont
  {Kramer}}, \bibinfo {author} {\bibfnamefont {C.}~\bibnamefont {Kreisbeck}}, \
  and\ \bibinfo {author} {\bibfnamefont {V.}~\bibnamefont {Krueckl}},\ }\href
  {http://stacks.iop.org/1402-4896/82/i=3/a=038101} {\bibfield  {journal}
  {\bibinfo  {journal} {Physica Scripta}\ }\textbf {\bibinfo {volume} {82}},\
  \bibinfo {pages} {038101} (\bibinfo {year} {2010})}\BibitemShut {NoStop}%
\bibitem [{\citenamefont {B\"uttiker}\ \emph {et~al.}(1993)\citenamefont
  {B\"uttiker}, \citenamefont {Thomas},\ and\ \citenamefont
  {PrÃÂªtre}}]{BUTTIKERPLA1993}%
  \BibitemOpen
  \bibfield  {author} {\bibinfo {author} {\bibfnamefont {M.}~\bibnamefont
  {B\"uttiker}}, \bibinfo {author} {\bibfnamefont {H.}~\bibnamefont {Thomas}},
  \ and\ \bibinfo {author} {\bibfnamefont {A.}~\bibnamefont {PrÃÂªtre}},\
  }\href {\doibase https://doi.org/10.1016/0375-9601(93)91193-9} {\bibfield
  {journal} {\bibinfo  {journal} {Physics Letters A}\ }\textbf {\bibinfo
  {volume} {180}},\ \bibinfo {pages} {364 } (\bibinfo {year}
  {1993})}\BibitemShut {NoStop}%
\bibitem [{\citenamefont {Mah\'e}\ \emph {et~al.}(2010)\citenamefont {Mah\'e},
  \citenamefont {Parmentier}, \citenamefont {Bocquillon}, \citenamefont
  {Berroir}, \citenamefont {Glattli}, \citenamefont {Kontos}, \citenamefont
  {Pla\ifmmode~\mbox{\c{c}}\else \c{c}\fi{}ais}, \citenamefont {F\`eve},
  \citenamefont {Cavanna},\ and\ \citenamefont {Jin}}]{MAHEPRB2010}%
  \BibitemOpen
  \bibfield  {author} {\bibinfo {author} {\bibfnamefont {A.}~\bibnamefont
  {Mah\'e}}, \bibinfo {author} {\bibfnamefont {F.~D.}\ \bibnamefont
  {Parmentier}}, \bibinfo {author} {\bibfnamefont {E.}~\bibnamefont
  {Bocquillon}}, \bibinfo {author} {\bibfnamefont {J.-M.}\ \bibnamefont
  {Berroir}}, \bibinfo {author} {\bibfnamefont {D.~C.}\ \bibnamefont
  {Glattli}}, \bibinfo {author} {\bibfnamefont {T.}~\bibnamefont {Kontos}},
  \bibinfo {author} {\bibfnamefont {B.}~\bibnamefont
  {Pla\ifmmode~\mbox{\c{c}}\else \c{c}\fi{}ais}}, \bibinfo {author}
  {\bibfnamefont {G.}~\bibnamefont {F\`eve}}, \bibinfo {author} {\bibfnamefont
  {A.}~\bibnamefont {Cavanna}}, \ and\ \bibinfo {author} {\bibfnamefont
  {Y.}~\bibnamefont {Jin}},\ }\href {\doibase 10.1103/PhysRevB.82.201309}
  {\bibfield  {journal} {\bibinfo  {journal} {Phys. Rev. B}\ }\textbf {\bibinfo
  {volume} {82}},\ \bibinfo {pages} {201309} (\bibinfo {year}
  {2010})}\BibitemShut {NoStop}%
\bibitem [{\citenamefont {Bocquillon}\ \emph {et~al.}(2013)\citenamefont
  {Bocquillon}, \citenamefont {Freulon}, \citenamefont {Berroir}, \citenamefont
  {Degiovanni}, \citenamefont {Pla{\c c}ais}, \citenamefont {Cavanna},
  \citenamefont {Jin},\ and\ \citenamefont {F{\`e}ve}}]{BOCQUILLONSCIENCE2013}%
  \BibitemOpen
  \bibfield  {author} {\bibinfo {author} {\bibfnamefont {E.}~\bibnamefont
  {Bocquillon}}, \bibinfo {author} {\bibfnamefont {V.}~\bibnamefont {Freulon}},
  \bibinfo {author} {\bibfnamefont {J.-M.}\ \bibnamefont {Berroir}}, \bibinfo
  {author} {\bibfnamefont {P.}~\bibnamefont {Degiovanni}}, \bibinfo {author}
  {\bibfnamefont {B.}~\bibnamefont {Pla{\c c}ais}}, \bibinfo {author}
  {\bibfnamefont {A.}~\bibnamefont {Cavanna}}, \bibinfo {author} {\bibfnamefont
  {Y.}~\bibnamefont {Jin}}, \ and\ \bibinfo {author} {\bibfnamefont
  {G.}~\bibnamefont {F{\`e}ve}},\ }\href {\doibase 10.1126/science.1232572}
  {\bibfield  {journal} {\bibinfo  {journal} {Science}\ }\textbf {\bibinfo
  {volume} {339}},\ \bibinfo {pages} {1054} (\bibinfo {year}
  {2013})}\BibitemShut {NoStop}%
\bibitem [{\citenamefont {Blumenthal}\ \emph {et~al.}(2007)\citenamefont
  {Blumenthal}, \citenamefont {Kaestner}, \citenamefont {Li}, \citenamefont
  {Giblin}, \citenamefont {Janssen}, \citenamefont {Pepper}, \citenamefont
  {Anderson}, \citenamefont {Jones},\ and\ \citenamefont
  {Ritchie}}]{BLUMENTHALNATPHYS201307}%
  \BibitemOpen
  \bibfield  {author} {\bibinfo {author} {\bibfnamefont {M.~D.}\ \bibnamefont
  {Blumenthal}}, \bibinfo {author} {\bibfnamefont {B.}~\bibnamefont
  {Kaestner}}, \bibinfo {author} {\bibfnamefont {L.}~\bibnamefont {Li}},
  \bibinfo {author} {\bibfnamefont {S.}~\bibnamefont {Giblin}}, \bibinfo
  {author} {\bibfnamefont {T.~J. B.~M.}\ \bibnamefont {Janssen}}, \bibinfo
  {author} {\bibfnamefont {M.}~\bibnamefont {Pepper}}, \bibinfo {author}
  {\bibfnamefont {D.}~\bibnamefont {Anderson}}, \bibinfo {author}
  {\bibfnamefont {G.}~\bibnamefont {Jones}}, \ and\ \bibinfo {author}
  {\bibfnamefont {D.~A.}\ \bibnamefont {Ritchie}},\ }\href {\doibase
  10.1038/nphys58} {\bibfield  {journal} {\bibinfo  {journal} {Nature Physics}\
  }\textbf {\bibinfo {volume} {3393}} (\bibinfo {year} {2007}),\
  10.1038/nphys58}\BibitemShut {NoStop}%
\bibitem [{\citenamefont {Dubois}\ \emph {et~al.}(2013)\citenamefont {Dubois},
  \citenamefont {Jullien}, \citenamefont {Portier}, \citenamefont {Roche},
  \citenamefont {Cavanna}, \citenamefont {Jin}, \citenamefont {Wegscheider},
  \citenamefont {Roulleau},\ and\ \citenamefont {Glattli}}]{DUBOISNATURE2013}%
  \BibitemOpen
  \bibfield  {author} {\bibinfo {author} {\bibfnamefont {J.}~\bibnamefont
  {Dubois}}, \bibinfo {author} {\bibfnamefont {T.}~\bibnamefont {Jullien}},
  \bibinfo {author} {\bibfnamefont {F.}~\bibnamefont {Portier}}, \bibinfo
  {author} {\bibfnamefont {P.}~\bibnamefont {Roche}}, \bibinfo {author}
  {\bibfnamefont {A.}~\bibnamefont {Cavanna}}, \bibinfo {author} {\bibfnamefont
  {Y.}~\bibnamefont {Jin}}, \bibinfo {author} {\bibfnamefont {W.}~\bibnamefont
  {Wegscheider}}, \bibinfo {author} {\bibfnamefont {P.}~\bibnamefont
  {Roulleau}}, \ and\ \bibinfo {author} {\bibfnamefont {D.~C.}\ \bibnamefont
  {Glattli}},\ }\href {\doibase 10.1038/nature12713} {\bibfield  {journal}
  {\bibinfo  {journal} {Nature}\ }\textbf {\bibinfo {volume} {502}},\ \bibinfo
  {pages} {659} (\bibinfo {year} {2013})}\BibitemShut {NoStop}%
\bibitem [{\citenamefont {Kataoka}\ \emph {et~al.}(2017)\citenamefont
  {Kataoka}, \citenamefont {Fletcher},\ and\ \citenamefont
  {Johnson}}]{KATAOKAPSS(2017)}%
  \BibitemOpen
  \bibfield  {author} {\bibinfo {author} {\bibfnamefont {M.}~\bibnamefont
  {Kataoka}}, \bibinfo {author} {\bibfnamefont {J.~D.}\ \bibnamefont
  {Fletcher}}, \ and\ \bibinfo {author} {\bibfnamefont {N.}~\bibnamefont
  {Johnson}},\ }\href {\doibase 10.1002/pssb.201600547} {\bibfield  {journal}
  {\bibinfo  {journal} {physica status solidi (b)}\ }\textbf {\bibinfo {volume}
  {254}},\ \bibinfo {pages} {1600547} (\bibinfo {year} {2017})}\BibitemShut
  {NoStop}%
\bibitem [{\citenamefont {Bauerle}\ \emph {et~al.}(2018)\citenamefont
  {Bauerle}, \citenamefont {Glattli}, \citenamefont {Meunier}, \citenamefont
  {Portier}, \citenamefont {Roche}, \citenamefont {Roulleau}, \citenamefont
  {Takada},\ and\ \citenamefont {Waintal}}]{Bauerle_RPP2018}%
  \BibitemOpen
  \bibfield  {author} {\bibinfo {author} {\bibfnamefont {C.}~\bibnamefont
  {Bauerle}}, \bibinfo {author} {\bibfnamefont {D.~C.}\ \bibnamefont
  {Glattli}}, \bibinfo {author} {\bibfnamefont {T.}~\bibnamefont {Meunier}},
  \bibinfo {author} {\bibfnamefont {F.}~\bibnamefont {Portier}}, \bibinfo
  {author} {\bibfnamefont {P.}~\bibnamefont {Roche}}, \bibinfo {author}
  {\bibfnamefont {P.}~\bibnamefont {Roulleau}}, \bibinfo {author}
  {\bibfnamefont {S.}~\bibnamefont {Takada}}, \ and\ \bibinfo {author}
  {\bibfnamefont {X.}~\bibnamefont {Waintal}},\ }\href
  {http://stacks.iop.org/0034-4885/81/i=5/a=056503} {\bibfield  {journal}
  {\bibinfo  {journal} {Reports on Progress in Physics}\ }\textbf {\bibinfo
  {volume} {81}},\ \bibinfo {pages} {056503} (\bibinfo {year}
  {2018})}\BibitemShut {NoStop}%
\bibitem [{\citenamefont {Ryu}\ \emph {et~al.}(2016)\citenamefont {Ryu},
  \citenamefont {Kataoka},\ and\ \citenamefont {Sim}}]{RYUPRL2016}%
  \BibitemOpen
  \bibfield  {author} {\bibinfo {author} {\bibfnamefont {S.}~\bibnamefont
  {Ryu}}, \bibinfo {author} {\bibfnamefont {M.}~\bibnamefont {Kataoka}}, \ and\
  \bibinfo {author} {\bibfnamefont {H.-S.}\ \bibnamefont {Sim}},\ }\href
  {\doibase 10.1103/PhysRevLett.117.146802} {\bibfield  {journal} {\bibinfo
  {journal} {Phys. Rev. Lett.}\ }\textbf {\bibinfo {volume} {117}},\ \bibinfo
  {pages} {146802} (\bibinfo {year} {2016})}\BibitemShut {NoStop}%
\bibitem [{\citenamefont {Ronetti}\ \emph {et~al.}(2020)\citenamefont
  {Ronetti}, \citenamefont {Carrega},\ and\ \citenamefont
  {Sassetti}}]{RonettiPRR2020}%
  \BibitemOpen
  \bibfield  {author} {\bibinfo {author} {\bibfnamefont {F.}~\bibnamefont
  {Ronetti}}, \bibinfo {author} {\bibfnamefont {M.}~\bibnamefont {Carrega}}, \
  and\ \bibinfo {author} {\bibfnamefont {M.}~\bibnamefont {Sassetti}},\ }\href
  {\doibase 10.1103/PhysRevResearch.2.013203} {\bibfield  {journal} {\bibinfo
  {journal} {Phys. Rev. Research}\ }\textbf {\bibinfo {volume} {2}},\ \bibinfo
  {pages} {013203} (\bibinfo {year} {2020})}\BibitemShut {NoStop}%
\bibitem [{\citenamefont {Keeling}\ \emph {et~al.}(2006)\citenamefont
  {Keeling}, \citenamefont {Klich},\ and\ \citenamefont
  {Levitov}}]{KEELINGPRL2006}%
  \BibitemOpen
  \bibfield  {author} {\bibinfo {author} {\bibfnamefont {J.}~\bibnamefont
  {Keeling}}, \bibinfo {author} {\bibfnamefont {I.}~\bibnamefont {Klich}}, \
  and\ \bibinfo {author} {\bibfnamefont {L.~S.}\ \bibnamefont {Levitov}},\
  }\href {\doibase 10.1103/PhysRevLett.97.116403} {\bibfield  {journal}
  {\bibinfo  {journal} {Phys. Rev. Lett.}\ }\textbf {\bibinfo {volume} {97}},\
  \bibinfo {pages} {116403} (\bibinfo {year} {2006})}\BibitemShut {NoStop}%
\end{thebibliography}%
\bibliographystyle{apsrev4-1}

\end{document}